\documentstyle[12pt,epsf]{article}
\newcommand {\m}{\mu}
\newcommand {\n}{\nu}
\newcommand {\pl}{\partial}
\newcommand {\p} {\phi}

\newcommand {\al}{\alpha}
\newcommand {\be}{\beta}
\newcommand {\ga}{\gamma}
\newcommand {\Ga}{\Gamma}

\newcommand {\ka}{\kappa}
\newcommand {\la}{\lambda}

\newcommand {\si}{\sigma}

\newcommand {\th}{\theta}

\newcommand {\om}{\omega}

\newcommand {\ep}{\epsilon}

\newcommand {\e} {\mbox{\rm e}}
\newcommand {\na}{\nabla}
\newcommand {\del}  {\delta}
\newcommand {\Del}  {\Delta}
\newcommand {\mn}{{\mu\nu}}
\newcommand {\ls}   {{\lambda\sigma}}
\newcommand {\ab}   {{\alpha\beta}}
\newcommand {\half}{ {\frac{1}{2}} }
\newcommand {\fourth} {\frac{1}{4} }
\newcommand {\sqg} {\sqrt{g}}

\newcommand {\Lcal}{{\cal L}}

\newcommand {\Dcal}{{\cal D}}

\newcommand {\Vvec}{{\vec V}}

\newcommand {\change} {\leftrightarrow}
\newcommand {\ra} {\rightarrow}
\newcommand {\pr}   {{\quad .}}
\newcommand {\com}  {{\quad ,}}
\newcommand {\q}    {\quad}
\newcommand {\qq}   {\quad\quad}
\newcommand {\qqq}   {\quad\quad\quad}

\newcommand {\lb}    {\linebreak}
\newcommand {\nl}    {\newline}
\newcommand {\nn}    {\nonumber}
\newcommand {\ul}    {\underline}
\newcommand {\vs}[1]  { \vspace*{#1 cm} }

\newcounter{eq}
\newcounter{sc}

\def\overleftrightarrow#1{\vbox{\ialign{##\crcr
 $\leftrightarrow$\crcr\noalign{\kern-1pt\nointerlineskip}
 $\hfil\displaystyle{#1}\hfil$\crcr}}}

\begin{document}
\title{    Weak Field Expansion of Gravity\ :\ 
           Graphs, Matrices and Topology
           \thanks{hep-th/9803238 , US-98-03}
                                 }
\author{  
          Shoichi ICHINOSE$^a$
          \thanks{ E-mail address:\ ichinose@u-shizuoka-ken.ac.jp}
          and Noriaki IKEDA$^b$
          \thanks{ E-mail address:\ nori@kurims.kyoto-u.ac.jp }\\
     $^a$ Department of Physics, University of Shizuoka,       \\
          Yada 52-1, Shizuoka 422-8526, Japan                       \\
     $^b$ Research Institute for Mathematical Sciences, \\
          Kyoto University, Kyoto 606-01,  Japan                  
        }
\date{  March, 1998 }
\maketitle
\setlength{\baselineskip}{0.54cm}
\begin{abstract}
We present some approaches to the perturbative analysis of the
classical and quantum gravity. First
we introduce a graphical representation for a global SO(n) tensor
$(\pl)^d h_\ab$, which generally appears 
in the weak field expansion around the flat space: 
$g_\mn=\del_\mn+h_\mn$. 
Making use of this representation, we explain
1) Generating function of graphs (Feynman diagram approach),
2) Adjacency matrix (Matrix approach),
3) Graphical classification in terms of "topology indices"
(Topology approach),
4) The Young tableau ( Symmetric group approach). 
We systematically construct the global SO(n) invariants.
How to show the independence and completeness of those invariants
is the main theme. We explain it
taking simple examples of 
$\pl\pl h-, \mbox{and}\ (\pl\pl h)^2-$ invariants in the text. 
The results are applied to the analysis of the independence of
general invariants and (the 
leading order of ) the Weyl anomalies of 
scalar-gravity theories in "diverse" dimensions
(2,4,6,8,10 dimensions).
\end{abstract}
\vs 1
\begin{itemize}
\item 
Key Words:\ weak field expansion, gravity, graphical representation, 
matrix representation, topology, SO(n)-invariants.
\item
PACS No.:\ 02.70.-c, 02.40.-k, 04.50.+h, 04.62.+v, 11.25.Db.
\end{itemize}
\section{Introduction}
\q Stimulated by the string theory, the higher ( than 4 ) dimensional
(super) gravity becomes more and more important as its low-energy 
(field theory) limit. In order to understand the coming new concept
involved in the string dynamics, it seems to become important to
analyze the higher dim gravity as the quantum field theory.
Generally, however, the higher dim gravity is technically difficult 
to treat with. We must deal with higher dim general invariants.
( For example, the conformal anomaly terms in $n$ dim are of the type:\ 
(Riemann tensor)$^{n/2}$. ) The present analysis aims at developing
some useful approaches in such analysis.

\q In $n$-dimensional Euclidean (Minkowski) flat space(-time), 
fields are classified as scalar, spinor, vector, tensor ... ,
by the
transformation property under the global SO(n) ( SO(n-1,1) ) transformation
of space(-time) coordinates.
\begin{eqnarray}
{x^\m}'=M^\m_{~\n}x^\n\com\q
M\in SO(n)
\com\q \label{intro.1}
\end{eqnarray}
where $M$ is a $n\times n$ matrix of SO(n)(SO(n-1,1)) 
\footnote{
Hereafter we take the Euclidean case for simplicity.
}.
As for the lower spin fields, the field theory is well defined classically
and quantumly. 

The general curved space is described by the general relativity which is based
on invariance under the general coordinate transformation.
Its infinitesimal form is written as
\begin{eqnarray}
{x^\m}'=x^\m-\ep^\m(x)\com\q |\ep|\ll 1\com\nn\\
\del g_\mn=g_{\m\la}\na_\n\ep^\la+g_{\n\la}\na_\m\ep^\la+O(\ep^2)
=\ep^\la \pl_\la g_\mn +g_{\m\la}\pl_\n\ep^\la+g_{\n\la}\pl_\m\ep^\la
+O(\ep^2)\com\q \label{intro.2}
\end{eqnarray}
where $\ep^\m$ is an infinitesimal local free
parameter. 
The general invariant which is 
composed of purely geometrical quantities and whose mass
dimension
\footnote{
We take the natural units: $c=\hbar=1$.
}
 is $(Mass)^2$, 
is uniquely given by Riemann scalar curvature $R$ 
defined by
\begin{eqnarray}
\Ga^\la_\mn=\half g^\ls (\pl_\m g_{\si\n}+\pl_\n g_{\si\m}-\pl_\si g_\mn)\com
R^\la_{~\m\n\si}=\pl_\n\Ga^\la_{\m\si}+\Ga^\la_{\tau\n}\Ga^\tau_{\m\si}-
\n\change\si\com\nn\\
R_\mn =R^\la_{~\mn\la}\com\q R=g^\mn R_\mn\com\q g=\mbox{det}g_\mn\pr
                                                      \label{intro.3}
\end{eqnarray}
It is well-known that the general relativity can be constructed purely
within the flat space first by introducing a symmetric second rank tensor
(Fierz-Pauli field) and then by requiring consistency in the field equation
in a perturbative way of the weak field \cite{EA}. 
In the present case, we can obtain the perturbed lagrangian simply by
the perturbation around the flat space.
\begin{eqnarray}
g_\mn=\del_\mn+h_\mn\com\q |h_\mn|\ll 1\pr
                                                      \label{intro.4}
\end{eqnarray}
Then the transformation (\ref{intro.2}) is expressed as
\begin{eqnarray}
\del h_\mn=\pl_\m\ep^\n+h_{\m\la}\pl_\n\ep^\la+\half \ep^\la \pl_\la h_\mn
+\m\change \n +O(\ep^2)\pr
                                                      \label{intro.5}
\end{eqnarray}
In the right-hand side (RHS), there appear 
$h^0$-order terms and $h^1$-order terms.
Therefore the general coordinate transformation (\ref{intro.5}) does 
{\it not}
preserve the weak-field ($h_\mn$) perturbation order.
(This is the reason why the result based on the lower-order
weak-field perturbation restores all higher-order terms
after the requirement of the general invariance.)
Riemann scalar curvature is also expanded as
\begin{eqnarray}
& R=\pl^2 h-\pl_\m\pl_\n h_\mn
 -h_\mn (\pl^2 h_\mn-2\pl_\la\pl_\m h_{\n\la}+\pl_\m\pl_\n h) & \nn\\
& +\half \pl_\m h_{\n\la}\cdot \pl_\n h_{\m\la}
-\frac{3}{4}\pl_\m h_{\n\la}\cdot \pl_\m h_{\n\la}
+\pl_\m h_{\m\la}\cdot \pl_\n h_{\n\la}                      & \nn\\
& -\pl_\m h_\mn\cdot\pl_\n h+\fourth\pl_\m h\cdot\pl_\m h
+O(h^3)\com\q                                          & \label{intro.6}
\end{eqnarray}
where $h\equiv h_{\m\m}$.
RHS is expanded to the infinite power
of $h_\mn$ due to the presence of the 'inverse' field of $g_\mn$,\ 
$g^\mn$, in (\ref{intro.3}).

It is explicitly checked that $R$, defined perturbatively by
the RHS of (\ref{intro.6}), transforms, under (\ref{intro.5}),
as a scalar $\del R(x)=\ep^\la(x)\pl_\la R(x)$, at the order of $O(h)$.
Because the general coordinate symmetry does not preserve the
the weak-field ($h_\mn$) perturbation order, we need
$O(h^2)$ terms in (\ref{intro.6}) in order to verify
$\del R(x)=\ep^\la(x)\pl_\la R(x)$, at the order of $O(h)$.
The first two terms of RHS of (\ref{intro.6}), 
$\pl^2 h$ and $\pl_\m\pl_\n h_\mn$, are two independent global
SO(n) invariants at the order O(h).
We may regard the weak field perturbation using (\ref{intro.4}) as a
sort of 'linear' representation of the general coordinate symmetry,
where all general invariant
quantities are generally expressed by the infinite series of power of $h_\mn$,
and there appears no 'inverse' fields.
One advantage of the linear representation is that the independence
of invariants, as a local function of $x^\m$, 
can be clearly shown because all quantities are written only by 
$h_\mn$ and its derivatives. 
We analyze some basic aspects of the weak-field expansion and develop
a useful graphical technique. 

It is a classical theme to obtain invariants with respect to
a group\cite{W}.
One established method to obtain $SO(n)$-invariants is to use the
representation theory of the symmetric group, or the Young tableaus
\cite{L}\cite{FKWC}.
It is reviewed in App.A, 
for the comparison with the present work.
We propose some new methods in this paper.
The basic idea is to express every $SO(n)$-invariant graphically (Sec.2)
and treat every procedure ( suffix contraction in the tensor product, 
classification of invariants, enumeration, e.t.c.)  in relation to
graphs. A part of these new methods is already used in Ref.\cite{II2}
where $\pl\pl h$-tensor is mainly considered. 
Their usefulness is confirmed by the complete classification of 
$(\pl\pl h)^3$-invariants. 
The present paper treats
a more general case: $h,\ \pl h,\ \pl\pl h,\ \pl\pl\pl h,\ \cdots$.

In Sec.2 the graphical representation of tensors 
and invariants is introduced. We start with
the field theory approach to the present problem in Sec.3.  
It is familiar in the perturbative field theory to use the
Feynman diagrams to express all expanded terms.  
The generating functional which generates
every $SO(n)$-invariant is given. 
There exist, in the graph theory\cite{H}, 
some matrix-representations to express a graph. We take
the {\it adjacency matrix} and apply it to the present problem in Sec.4.
All graphs of invariants is systematically listed
using the graph topology in Sec.5. 
In Sec.6, the completeness of the graph enumeration is shown
from the viewpoint of the suffix-permutation symmetry.
In order to identify every graph succinctly, we introduce
a set of {\it indices} in Sec.7. It is explicitly shown that 
every $(\pl\pl h)^2$-invariant is identified by five indices. 
In Sec.8, we explain how to read these indices from the adjacency
matrices. We apply the present results to some gravitational problem
in Sec.9. Finally we conclude in Sec.10. Four appendices are prepared
to complement the text. In App.A, we review the Young tableau approach
in relation to the present problem. Some examples of 
adjacency matrices are given
in App.B. In App.C we explain some indices in detail. The gauge-fixing
condition and  the corresponding graphical rule are explained in App.D.

\section{Graphical Representation of $\pl^d h$-tensors and invariants}
We treat a general invariant made from any contraction of 
the following quantity:
\begin{eqnarray}
\label{Rep.1}
(\partial_{\mu_1}\partial_{\mu_2}\cdots \partial_{\mu_{d_1}} 
h_{\lambda_1\rho_1})
(\partial_{\n_1}\partial_{\n_2}\cdots \partial_{\n_{d_2}}
h_{\lambda_2\rho_2})
\cdots
(\partial_{\tau_1}\partial_{\tau_2}\cdots \partial_{\tau_{d_B}}
h_{\lambda_B\rho_B})\ .
\end{eqnarray}
We define two basic indices of the invariant:\  
the mass dimension(= No. of differentials), $D$, and the number of 
$h_{\lambda\rho}$(=No. of "bonds" defined below), $B$.
If $D$ and $B$ are given, we can define another index, 
the {\it partition} of $D$, $(d_1, d_2, d_3, \cdots, d_B)$,
corresponding to a  division of differentiations.
The partition satisfies the following relation.
\begin{eqnarray}
\label{Rep.2}
D = \sum_{k=1}^B d_k\com\q d_k:\ 0,1,2,3,\cdots\com
\end{eqnarray}
where $d_k$ is the number of differentiations of 
$h_{\lambda_k\rho_k}$.

\q 
We graphically represent a global $SO(n)$ tensor,
\begin{eqnarray}
\label{Rep.3}
\partial_{\mu_1}\partial_{\mu_2}\cdots \partial_{\mu_k} h_{\lambda\rho}\pr
\end{eqnarray}
as in Fig.1\cite{II2}.
\begin{figure}
  \centerline{\epsfysize=4cm\epsfbox{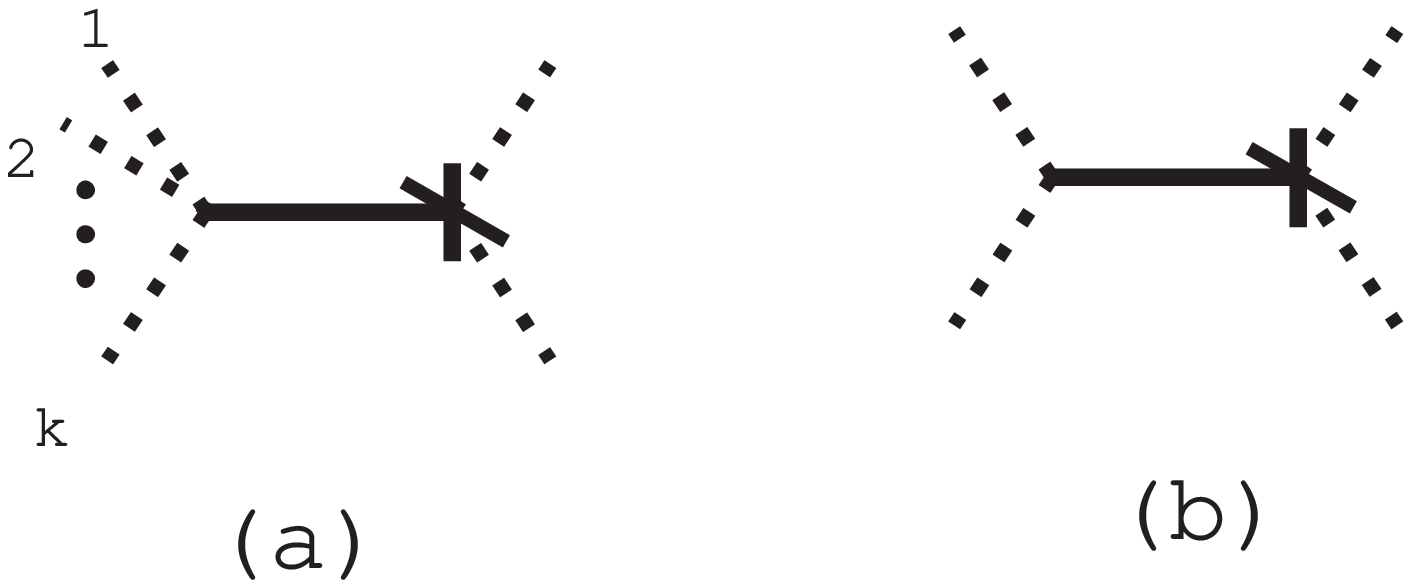}}
\begin{center}
{Fig.1\ (a)\ (k+2)-tensor (\ref{Rep.3});
\ (b)\ 4-tensor $\pl_\m\pl_\n h_\ab$}
\end{center}
\end{figure}
The graph respects all suffix-permutation
symmetries of 
$\partial_{\mu_1}\partial_{\mu_2}\cdots \partial_{\mu_k} h_{\lambda\rho}$:
\begin{enumerate}
\item totally symmetric with repect to $(\mu_1,\mu_2,\cdots,\mu_k)$,  
\item symmetric with repect to $\la$ and $\rho$. 
\end{enumerate}
Let us introduce some definitions.
\begin{description}
\item[Def 1]\cite{II2}\ 
We call dotted lines  {\it suffix-lines}, a rigid line a {\it bond},
a vertex with a crossing mark a {\it h-vertex} and that without it 
a {\it der-vertex}. We also specifically call the last one {\it d$^k$-vertex}
where $k$ is the number of suffix-lines 
which the crossing-mark-less vertex has.
We often use {\it dd-vertex} instead of {\it d$^2$-vertex}.
\end{description}
\begin{description}
\item[Def 2]\cite{II2}\ 
The suffix {\it contraction} is graphically expressed 
by connecting the two corresponding 
suffix-lines. 
\end{description}
For example, 2nd rank tensors (2-tensors)
\footnote{
We call k-th rank tensor {\it k-tensor} hereafter.
}
\ :\ 
$\pl^2h_\ab\ ,\ \pl_\m\pl_\n h_{\al\al}\ ,\ \pl_\m\pl_\be h_\ab\ $
, which are made from the Fig.1b 
by connecting two suffix-lines, 
are expressed as in Fig.2.
\begin{figure}
  \centerline{\epsfysize=4cm\epsfbox{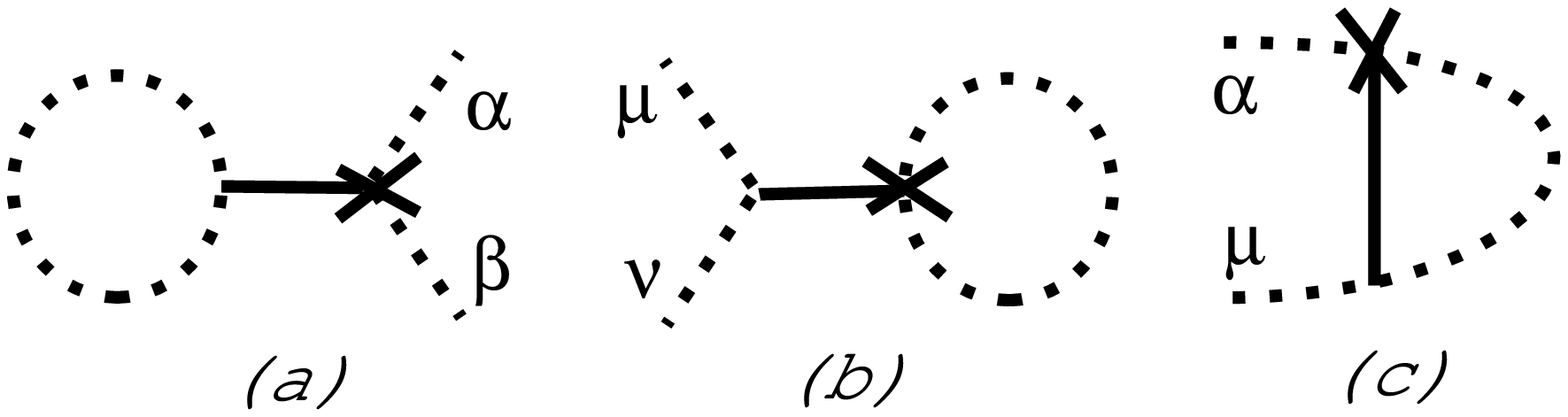}}
   \begin{center}
Fig.2
\ 2-tensors of 
$\pl^2h_\ab\ ,\ \pl_\m\pl_\n h_{\al\al}\ $ and $\pl_\m\pl_\be h_\ab\ $
   \end{center}
\end{figure}   
Two independent  invariants (0-tensors)\ :\ 
$P\equiv \pl_\m\pl_\m h_{\al\al},\ Q\equiv \pl_\al\pl_\be h_{\ab}\ $
, which are made from Fig.2 by connecting the remaining two suffix-lines, are
expressed as in Fig.3.
\begin{figure}
\centerline{\epsfysize=4cm\epsfbox{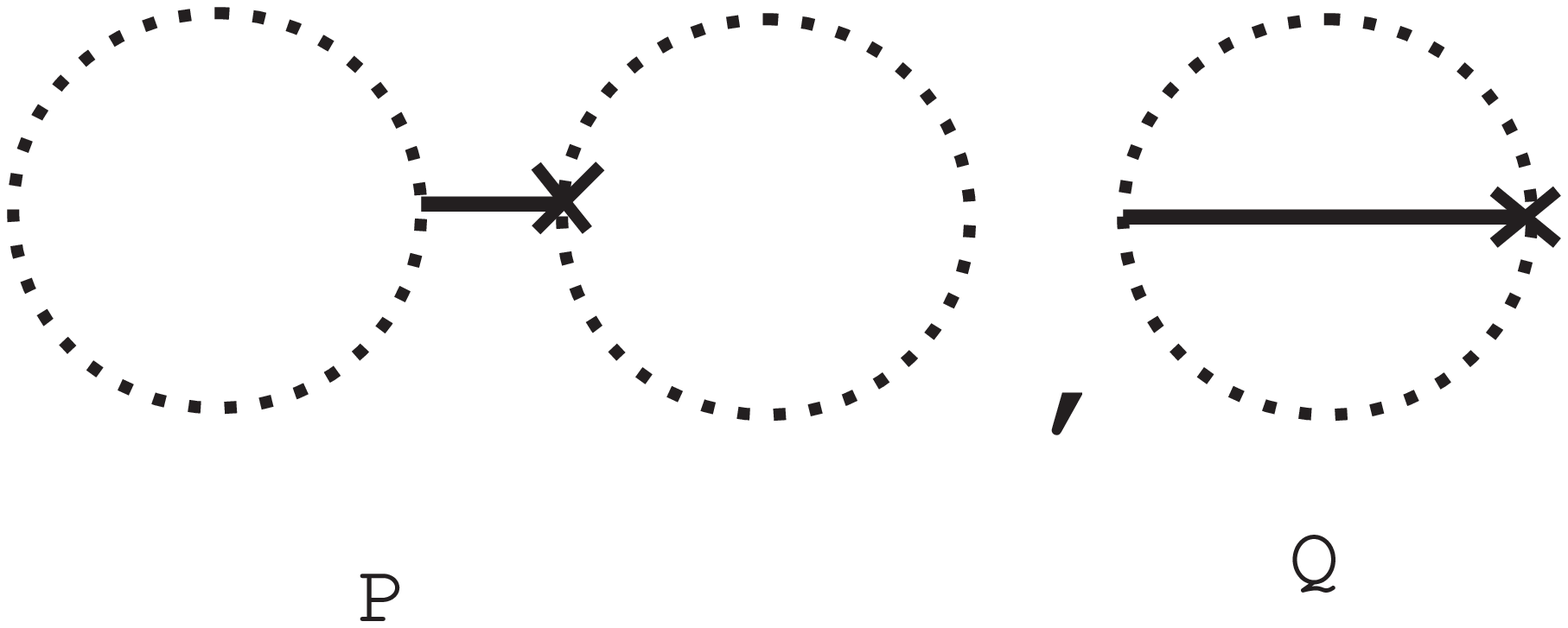}}
   \begin{center}
Fig.3\ Invariants of 
$P\equiv \pl_\m\pl_\m h_{\al\al}$\ and $Q\equiv \pl_\al\pl_\be h_{\ab}\ $.
   \end{center}
\end{figure}   
P and Q are all possible invariants  of $\pl\pl h$-type.
All suffix-lines of Fig.3 are closed. We easily see the following lemma
is valid.
\begin{description}
\item[Lemma 1]\cite{II2}\ 
Generally all suffix-lines of invariants are  {\it closed}.
We call a closed suffix-line a {\it suffix-loop}.
\end{description}

If we neglect degeneracies due to
symmetries among suffixes. 
the number of different contractions of 
the general tensor (\ref{Rep.1}) to make invariants
is given  as
\begin{eqnarray}
(N-1)(N-3)\cdots3\cdot1
=\frac{(N)!}{2^{\frac{N}{2}}(\frac{N}{2})!}\com\q N\equiv D+2B\com
\label{sumweight}
\end{eqnarray}
This relation
turns out to be important to show the enumeration of all graphs
with no missing graphs.

\section{Generating Functional for Graphs:\ Feynman Diagram Approach}

We can construct the generating functional for all possible graphs.
It is an application of the field theoretic approach 
to our graphical representations.
Let us consider
the following Lagrangian in $2$ space-dimension
\footnote{
The space-dimension is taken to be 2 in order to
obtain the mass-dimensions of the couplings, (\ref{conc.3}), 
without
introducing any additional mass parameters. Note that
the present purpose is the graph classification.
There the important thing is the topological structure
of graphs. 
The space-dimension of the field theory is irrelevant. 
}
. 
\begin{eqnarray}
\Lcal[\p,\om_1,\om_2]=\Lcal_0+\Lcal_I\com\nn\\
\Lcal_0=\half\p^2+\om_1\om_2\com\q
\Lcal_I[\p,\om_1,\om_2]
=\om_1~\sum_{i=0}^{\infty}g_i\p^i+\la\p^2\om_2\pr
     \label{conc.1}
\end{eqnarray}
The dotted lines in Sec.2 are
represented by  $\phi\phi$ propagators and 
the rigid lines by $\omega_1\omega_2$ propagators.
$\la$-interaction term describes the $h$-vertex and
$g_i$-interaction term describes the $d^i$-vertex
See Fig.4.
\begin{figure}
\centerline{\epsfysize=10cm\epsfbox{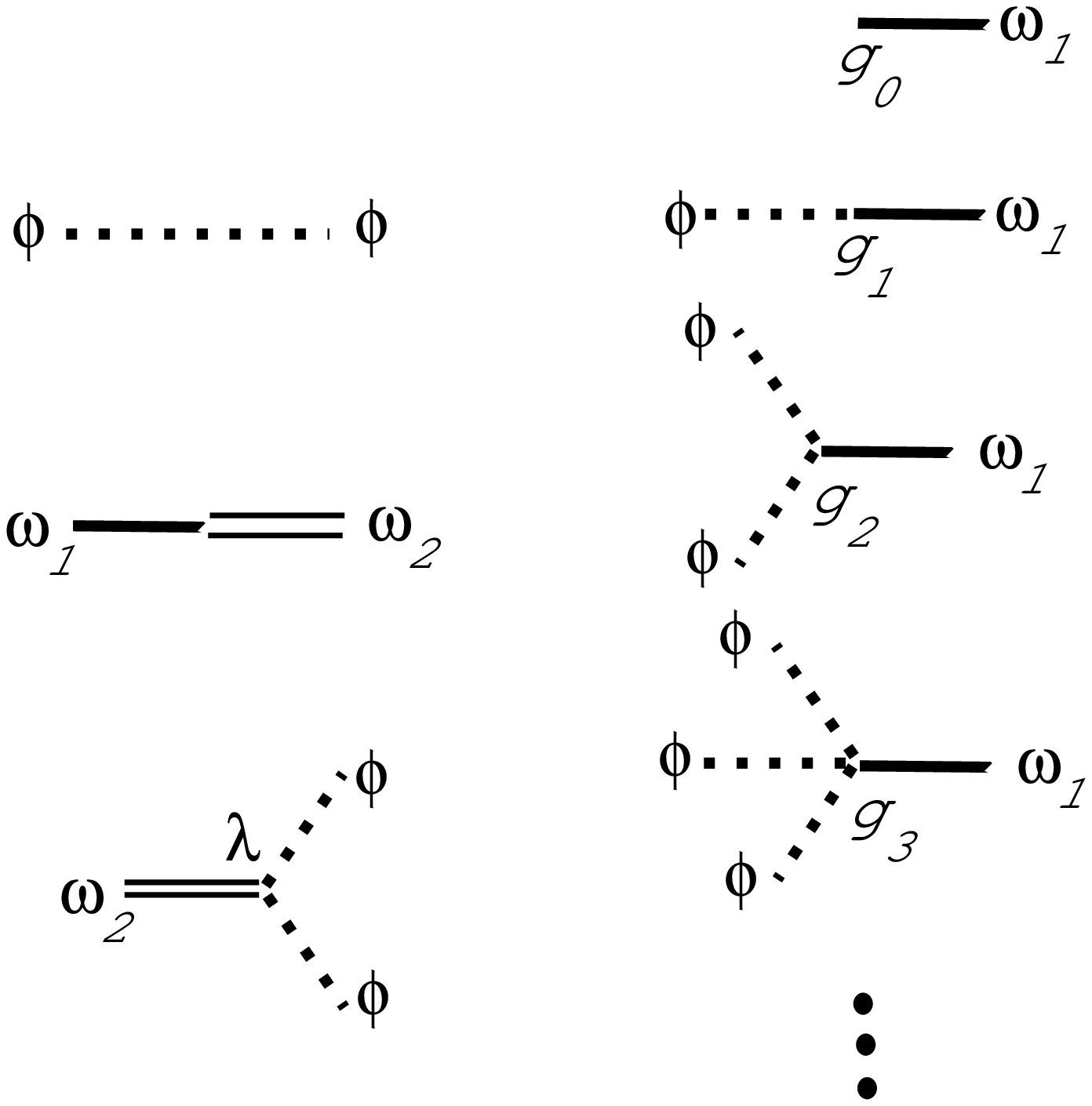}}
   \begin{center}
Fig.4\ Feynman rule of the Lagrangian (\ref{conc.1}).
   \end{center}
\end{figure}
We assign mass-dimension  as follows.
\begin{eqnarray}
[\Lcal]=M^2\com\q
[\p]=M\com\q
[\om_1]=M^2\com\q
[\om_2]=M^0\pr
     \label{conc.2}
\end{eqnarray}
Then we obtain
\begin{eqnarray}
[g_i]=M^{-i}\com\q
[\la]=M^0\pr
     \label{conc.3}
\end{eqnarray}

The generating functional of all graphs ( $SO(n)$-invariants,
$SO(n)$-tensors ) is given by
\begin{eqnarray}
& W[J,K_1,K_2]=\e^{\Ga[J,K_1,K_2]} \nn &\\
& =\int\Dcal\p\Dcal\om_1\Dcal\om_2\exp \left[
\int d^2x (\Lcal[\p,\om_1,\om_2]+J\p+K_1\om_1+K_2\om_2) \right]  
                                      \nn &\\
& =\sum_{r=0}^{\infty}\frac{1}{r!}\left[
\int d^2x \Lcal_I(\frac{\del}{\del J(x)},\frac{\del}{\del K_1(x)},
\frac{\del}{\del K_2(x)})       \right]^r
\exp \int d^2x (-\half J(x)J(x)-K_1(x)K_2(x) )\ .\nn &\\
                                     \label{conc.4}
\end{eqnarray}
All graphs of connected $k$-tensors appear in the $k$-point
Green function.
\begin{eqnarray}
\frac{1}{k!}\left.
\frac{\del}{\del J(x_1)}\frac{\del}{\del J(x_2)}\cdots
\frac{\del}{\del J(x_k)}\Ga[J,K_1,K_2]\right|_{J=0,K_1=0,K_2=0}\pr
                                     \label{conc.5}
\end{eqnarray}
In particular all $SO(n)$-invariants appear in the $k=0$ case above.
\begin{eqnarray}
\left.
\Ga[J,K_1,K_2]\right|_{J=0,K_1=0,K_2=0}\pr
                                     \label{conc.6}
\end{eqnarray}
They are given by perturbation with respect to (w.r.t) 
the couplings ($g_0.g_1,g_2,\cdots; \la$) in $\Lcal_I$.
For example, $(\pl\pl h)^s$-invariants ($s=1,2,\cdots$) are
given by $(g_2\la)^s$-terms ($r=2s$) in (\ref{conc.6}). 
From the coupling-dependence, we can read the mass-dimension of
each graph. They show the inverse mass-dimensions of corresponding graphs.
For example, from the relations   
$[(g_2\la)^s]=M^{-2s}$, 
$[g_4\cdot g_2\cdot \la^2]=M^{-6}$ and 
$[g_3\cdot g_3\cdot \la^2]=M^{-6}$ we see
$(\pl\pl h)^s, \pl^4h\cdot\pl^2h$ and $\pl^3h\cdot\pl^3h$-invariants
have the dimensions of $M^{2s},M^6,M^6$ respectively. 
The coefficient in front of each expanded term are related
with the {\it weight} of the corresponding graph which
will be soon defined.

We summarize this section. 
We can obtain all possible graphs by
enumerating all Feynman diagrams of the lagrangian (\ref{conc.1}).
The vacuum polarization graphs, in which there is no external line,
correspond to $SO(n)$-invariants.
The diagrams in which all external lines are dotted lines, $\phi\phi$
propagators, are $SO(n)$-tensors.
There is no corresponding tensorial objects 
for the diagrams in which external lines
include rigid lines, $\omega_1\omega_2$ propagators.
It is the advantage of this approach that the all invariants
can be generated from the compact formulae (\ref{conc.4}) and (\ref{conc.6}).


\section{Adjacency Matrix}
In Sec.2, we have defined the graphical representation of
an invariant. It is well known in the graph theory\cite{H}
that a graph can be represented by a matrix defined as follows.
The speciality of the present case is the "binary" structure
\footnote{
The general tensor (\ref{Rep.3}) is made of two parts:\ the derivative 
operator part and the weak-field ($h_{\la\rho}$) part.
}
of the tensor (\ref{Rep.3}) and its graph (Fig.1a). 
\begin{description}
\item[Def 3]\
We name the vertices of the graph of (\ref{Rep.1})
$v_1,v_{\bar 1};v_2,v_{\bar 2};\cdots ;v_B,v_{\bar B}$
as shown in Fig.5.
Then we define the {\it adjacency matrix} 
$A=[a_{IJ}];\ I,J=1,{\bar 1},2,{\bar 2},\cdots,B,{\bar B}$
as follows.\nl
i) $I\neq J$\nl
If $v_I$ is connected with $v_J$ by $k$ dotted lines (see Fig.6),
\begin{eqnarray}
a_{IJ}=a_{JI}=k.
\label{ad.1}
\end{eqnarray}
ii) $I=J$\nl
If $v_I$ is connected with itself by $l$ dotted lines (see Fig.7)
\begin{eqnarray}
a_{II}=2l.
\label{ad.2}
\end{eqnarray}
\end{description}
\begin{figure}
\centerline{\epsfysize=8cm\epsfbox{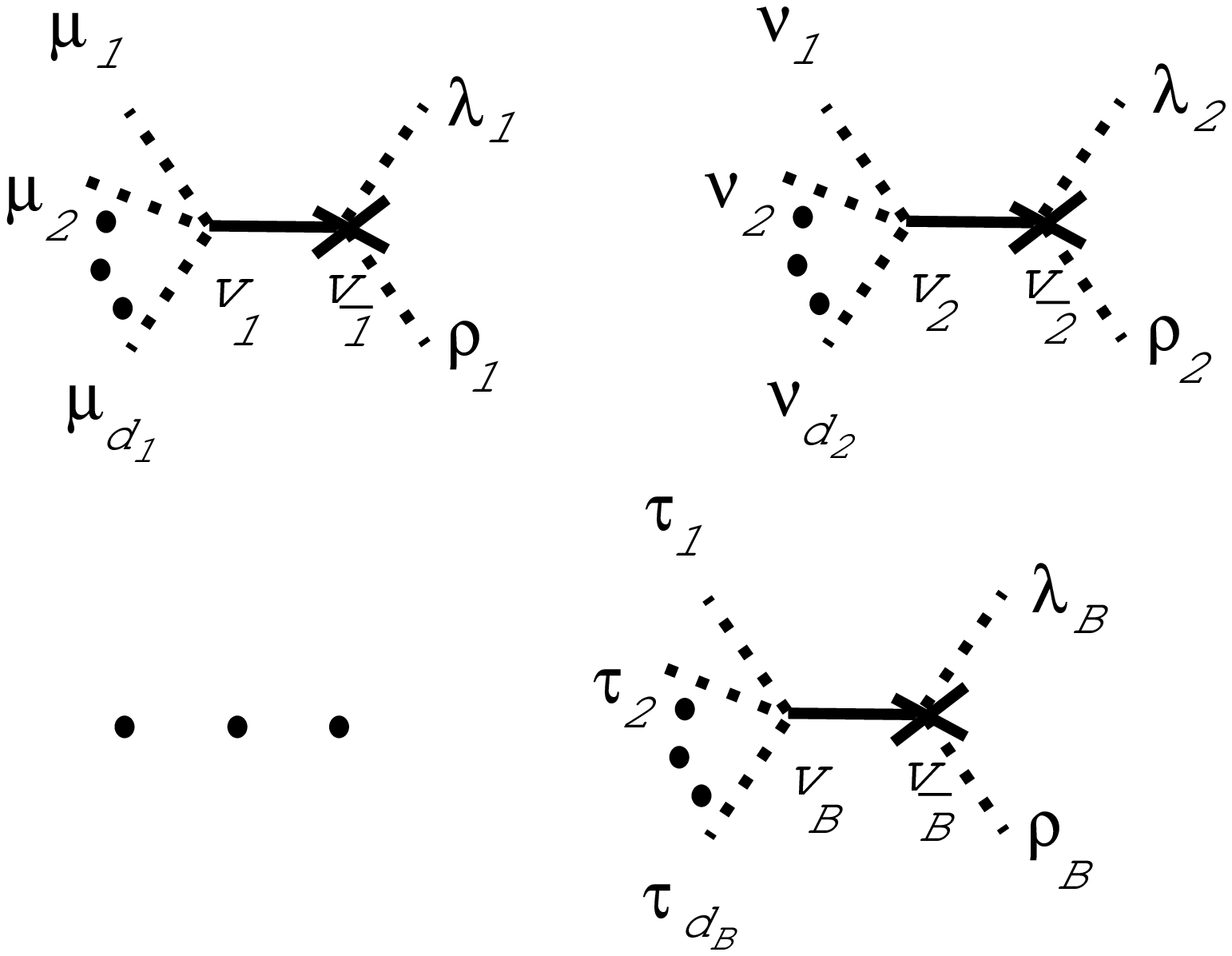}}
   \begin{center}
Fig.5\ Graph of (\ref{Rep.1}).
   \end{center}
\end{figure}
\begin{figure}
\centerline{\epsfysize=4cm\epsfbox{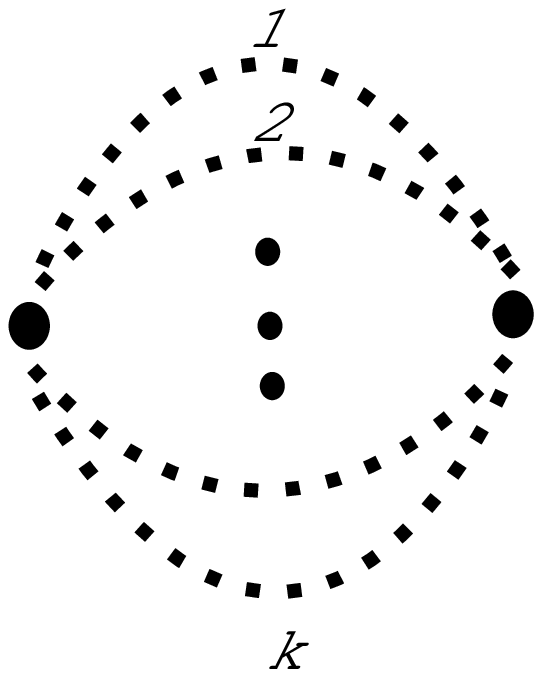}}
   \begin{center}
Fig.6\ Two vertices are connected by $k$ dotted lines. $\bullet$-vertex
represents a der-vertex or a h-vertex.
   \end{center}
\end{figure}
\begin{figure}
\centerline{\epsfysize=4cm\epsfbox{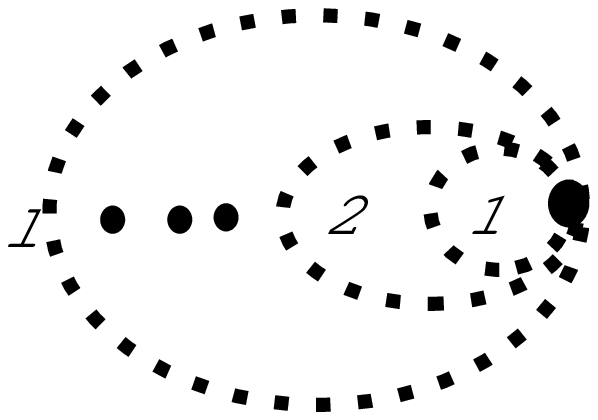}}
   \begin{center}
Fig.7\ A vertex is connected with itself by $l$ dotted lines. 
$\bullet$-vertex represents a der-vertex or a h-vertex.
   \end{center}
\end{figure}
Then we see the adjacency matrix $A$ satisfies the following
properties.
\begin{eqnarray}
a_{IJ}=a_{JI},(\mbox{symmetric})       \pr\nn\\
a_{IJ}=\left\{ \begin{array}{ll}
               0,1,2,\cdots (\mbox{non-negative integers}) & \mbox{for  } I\neq J\\
			   0,2,4,\cdots (\mbox{non-negative even integers})& \mbox{for  } I=J
                \end{array}
		\right.		                   \pr\nn\\
\sum_I a_{IJ}=\left\{ 
                     \begin{array}{lll}
               d_j & \mbox{for } J=j        &    \\
			       &                       & j=1,2,\cdots B\\
			   2   & \mbox{for } J={\bar j} &
                      \end{array}
		        \right.		            \pr
\label{ad.3}
\end{eqnarray}
The following relations can be derived
from the relations (\ref{ad.3}) and the condition (\ref{Rep.2}).
\begin{eqnarray}
\sum_J a_{IJ}=\left\{ 
                     \begin{array}{lll}
               d_i & \mbox{for } I=i        &    \\
			       &                       & i=1,2,\cdots B\\
			   2   & \mbox{for } I={\bar i} &
                      \end{array}
		        \right.		            \com\q       			
\sum_{IJ}a_{IJ}=D+2B\pr 		
\label{ad.3b}
\end{eqnarray}
When we will soon take the relations (\ref{ad.3}) as {\it necessary } conditions of the adjacency matrix
in its calculation algorithm (Step M2 below), the relations (\ref{ad.3b}) are automatically satisfied.
Because we have the freedom of naming the vertices 
$v_1,v_{\bar 1};v_2,v_{\bar 2};\cdots ;v_B,v_{\bar B}$, 
two matrices $A$ and $A'$, which are related by a $2B\times 2B$
permutation matrix $P$ in the following way, must be identified
($A\sim A'$). 
\begin{eqnarray}
A'\sim A\leftrightarrow A'=P^TAP\com      \nn\\
P=\left( \begin{array}{cc}
             1&0 \\
			 0&1
		  \end{array}
  \right)\bigotimes \hat{P}\com\q  \hat{P}\in \mbox{Permutation of }(1,2,\cdots,B)
                                                    \com
\label{ad.4}
\end{eqnarray}
\footnote{
In a specific case such as $d_i=d_{i'}\ (i\neq i')$,
$A'$ coincides with $A$ for some $P$'s due to 
the presence of some same-type 
terms among $B$ terms:\  \nl
$\pl^{d_1}h_{\la_1\rho_1},\pl^{d_2}h_{\la_2\rho_2},
\cdots, \pl^{d_B}h_{\la_B\rho_B}$. 
}.
With the above identification 
an adjacency matrix $A$ can be one-to-one with a graph.

Now we summarize this section by giving the algorithm to list 
all independent invariants for a given $D$(mass dimension) and
$B$(weak-field perturbation order, no. of bonds or $h$'s).
\begin{description}
\item[Step M1]\ Find all possible partitions $(d_1, d_2, d_3, \cdots, d_B)$
by solving (\ref{Rep.2}).
\item[Step M2]\ Find all possible adjacency matrices by solving
(\ref{ad.3}).
\item[Step M3]\ We introduce the equivalence relation (\ref{ad.4}) among the adjacency matrices.
\end{description}

As a simple example, we write the adjacency matrices of $P$ and $Q$\  
which are all possible $\pl\pl h$-invariants. 
\begin{eqnarray}
\label{Mddh}
&
P = \pl_\m\pl_\m h_{\al\al} =\ \mbox{Graph P in Fig.3}\simeq
\left[
\matrix{
2 & 0  \cr
0 & 2  \cr
}
\right]                                  \com & \nn\\
&
Q = \pl_\al\pl_\be h_{\al\be} =\ \mbox{Graph Q in Fig.3}\simeq
\left[
\matrix{
0 & 2  \cr
2 & 0  \cr
}
\right]\com
&
\end{eqnarray}
where $\simeq$ means that adjacency matrices are 
generally understood
up to the freedom of (\ref{ad.4}).
As the second example, we list all independent
$(\pl\pl h)^2$-invariants using the adjacency matrices. 
$D=4$ and $B=2$,
and the partition  is $(d_1,d_2)=(2, 2)$.
From the Step M1 to M3,
we can obtain the following 13 independent representatives 
of adjacency matrices:
\begin{eqnarray}
\label{Mddhddh}
&&
PP = (\pl^2 h_{\la\la})^2 \simeq
\left[
\matrix{
2 & 0 & 0 & 0 \cr
0 & 2 & 0 & 0 \cr
0 & 0 & 2 & 0 \cr
0 & 0 & 0 & 2 \cr
}
\right],
PQ = \pl^2 h_{\la\la}\cdot\pl_\m\pl_\n h_{\mn}
\simeq
\left[
\matrix{
2 & 0 & 0 & 0 \cr
0 & 2 & 0 & 0 \cr
0 & 0 & 0 & 2 \cr
0 & 0 & 2 & 0 \cr
}
\right],
\nonumber \\
&& C2 = \pl^2 h_{\mn}\cdot\pl^2 h_{\mn} \simeq
\left[
\matrix{
2 & 0 & 0 & 0 \cr
0 & 0 & 0 & 2 \cr
0 & 0 & 2 & 0 \cr
0 & 2 & 0 & 0 \cr
}
\right],
C3 = \pl_\m\pl_\n h_{\la\la}\cdot\pl^2 h_{\mn}
\simeq
\left[
\matrix{
2 & 0 & 0 & 0 \cr
0 & 0 & 2 & 0 \cr
0 & 2 & 0 & 0 \cr
0 & 0 & 0 & 2 \cr
}
\right],
\nonumber \\
&&
C1 
= \pl_\m\pl_\n h_{\la\la}\cdot\pl_\m\pl_\n h_{\si\si}
\simeq
\left[
\matrix{
0 & 0 & 2 & 0 \cr
0 & 2 & 0 & 0 \cr
2 & 0 & 0 & 0 \cr
0 & 0 & 0 & 2 \cr
}
\right],
B2 
= \pl^2 h_{\la\n}\cdot\pl_\la\pl_\m h_{\mn} \simeq
\left[
\matrix{
2 & 0 & 0 & 0 \cr
0 & 0 & 1 & 1 \cr
0 & 1 & 0 & 1 \cr
0 & 1 & 1 & 0 \cr
}
\right],
\nonumber \\
&& 
B1 = \pl_\n\pl_\la h_{\si\si}\cdot\pl_\la\pl_\m h_{\mn}
\simeq \left[
\matrix{
0 & 0 & 1 & 1 \cr
0 & 2 & 0 & 0 \cr
1 & 0 & 0 & 1 \cr
1 & 0 & 1 & 0 \cr
}
\right],
QQ = (\pl_\m\pl_\n h_{\mn})^2 \simeq
\left[
\matrix{
0 & 2 & 0 & 0 \cr
2 & 0 & 0 & 0 \cr
0 & 0 & 0 & 2 \cr
0 & 0 & 2 & 0 \cr
}
\right],
\nonumber \\
&&
A2 = \pl_\si\pl_\la h_{\la\m}\cdot\pl_\si\pl_\n h_{\mn}
\simeq
\left[
\matrix{
0 & 1 & 1 & 0 \cr
1 & 0 & 0 & 1 \cr
1 & 0 & 0 & 1 \cr
0 & 1 & 1 & 0 \cr
}
\right],
A3 = \pl_\si\pl_\la h_{\la\m}\cdot\pl_\m\pl_\n h_{\n\si}
\simeq
\left[
\matrix{
0 & 1 & 0 & 1 \cr
1 & 0 & 1 & 0 \cr
0 & 1 & 0 & 1 \cr
1 & 0 & 1 & 0 \cr
}
\right],
\nonumber \\
&& 
B4 = \pl_\m\pl_\n h_{\la\si}\cdot\pl_\la\pl_\si h_{\mn} \simeq
\left[
\matrix{
0 & 0 & 2 & 0 \cr
0 & 0 & 0 & 2 \cr
2 & 0 & 0 & 0 \cr
0 & 2 & 0 & 0 \cr
}
\right],
B3 = \pl_\m\pl_\n h_{\la\si}\cdot\pl_\m\pl_\n h_{\ls} \simeq
\left[
\matrix{
0 & 0 & 0 & 2 \cr
0 & 0 & 2 & 0 \cr
0 & 2 & 0 & 0 \cr
2 & 0 & 0 & 0 \cr
}
\right],
\nonumber \\
&&
A1 = \pl_\si\pl_\la h_\mn\cdot\pl_\si\pl_\n h_{\m\la} \simeq
\left[
\matrix{
0 & 0 & 1 & 1 \cr
0 & 0 & 1 & 1 \cr
1 & 1 & 0 & 0 \cr
1 & 1 & 0 & 0 \cr
}
\right],
\end{eqnarray}
where we have named the 13 invariants as above\cite{II2}
and they are used in the following sections.
The corresponding 13 graphs are  given in the next section.
In Appendix B, we list the adjacency matrices for 
another types of invariants:\ $(\partial \partial \partial h)^2$ and   
$(\partial \partial \partial \partial h \partial \partial h)$.  
They appear, for example, in the conformal anomaly calculation
in the 6 dim gravity-matter theory\cite{II98BNL}.

The advantages of the adjacency matrix representation are as follows:\ 
1) We can treat all invariants {\it without reference to graphs}, 
which allow us to do analysis in the algebraic way.
This representation is powerful in some analysis of 
general properties of graphs\cite{N};\ 
2) We can systematically treat invariants of 
wide-range types (cf. the graphical treatment in Sec.5).
On the other hand its disadvantage is the practical difficulty
of Step M3 in the algorithm. We will propose a way to resolve this point, 
in Sec.8, making use of the "topological" indices ( introduced in Sec.7).  

In Sec.3 and 4, we have presented two ways ( Feynman diagram and 
the adjacency matrix )
to cope with all invariants. Each of them is a closed formalism by itself. 
In the next section we present another
approach which is powerful in the practical calculation.

\section{
Graph Topology and List of All Invariants
}
\q From the viewpoint of the graph analysis, we present
a new approach to treat invariants. For simplicity we focus
on $(\pl\pl h)^s$-invariants.
This type terms typically appear in the weak-field expansion of
"products" of $s$ Riemann tensors. 
As examples of $(\pl\pl h)^2$-tensors, 
we have the representations of Fig.8
for $\pl_\m\pl_\n h_\ab \pl_\m\pl_\n h_{\ga\del}$ and
$\pl_\m\pl_\n h_\ab \pl_\n\pl_\la h_{\la\be}$.
\begin{figure}
\centerline{\epsfysize=4cm\epsfbox{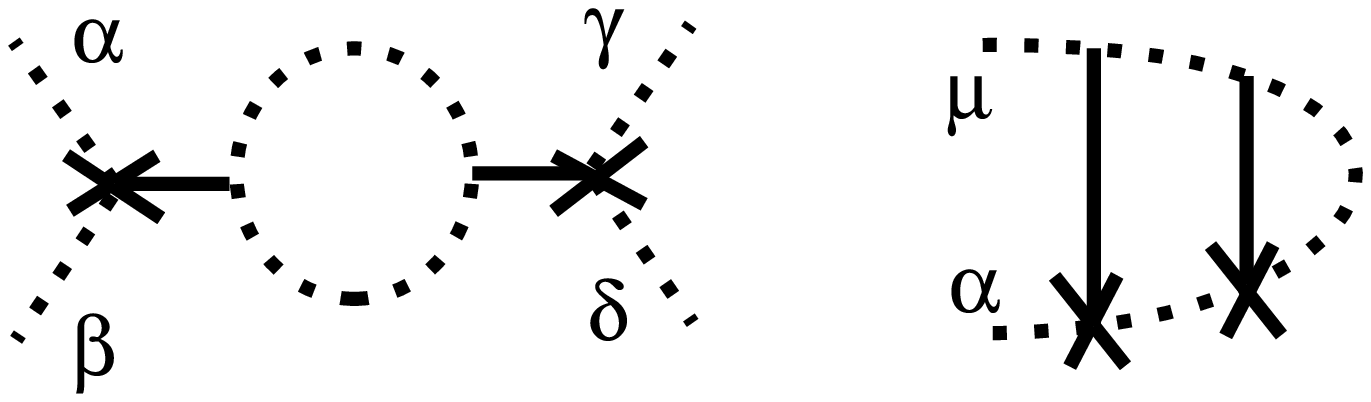}}
   \begin{center}
Fig.8\ 
Graphical Representations of 
$\pl_\m\pl_\n h_\ab \cdot\pl_\m\pl_\n h_{\ga\del}$ and
$\pl_\m\pl_\n h_\ab \cdot\pl_\n\pl_\la h_{\la\be}$.
   \end{center}
\end{figure}   
Let us state a lemma on a general SO(n)-invariant made of $s$ 
$\pl\pl h$-tensors.
\begin{description}
\item[Lemma 2]\cite{II2}\ 
Let a general $(\pl\pl h)^s$-invariant ($s=1,2,\cdots$) has $l$ suffix-loops.
Let
each loop have $v_i$ h-vertices and $w_i$ dd-vertices ($i=1,2,\cdots, l-1,l$).
We have the following {\it necessary} conditions for $s,l,v_i\ \mbox{and } w_i$.
\begin{eqnarray}
\sum_{i=1}^{l}v_i=s\com\q \sum_{i=1}^{l}w_i=s\com
\q v_i+w_i\geq 1\com\label{ddh2.1}\\
v_i\ ,\ w_i\ =0,1,2,\cdots\q ,\q l=1,2,3,\cdots,2s-1,2s\pr\nn
\end{eqnarray}
Here we may ignore the order of the elements\lb
 in a set
$\left\{
\left(\begin{array}{c}  v_i \\ w_i  \end{array}\right)\ ;\ 
i=1,2,\cdots,l-1,l
\right\}$
because the order can be arbitrarily changed by renumbering the suffix-loops.
\end{description}
This Lemma can be used to list all possible invariants. 

We consider the $s=2$ case ($(\pl\pl h)^2$-invariants) 
in order to show how to use Lemma 2 for enumerating all invariants.
\flushleft{(i) $l=1$}

\q For this case, we have 
\begin{eqnarray}
\left(\begin{array}{c}  v_1 \\ w_1  \end{array}\right)
=
\left(\begin{array}{c}  2 \\ 2  \end{array}\right)
\label{ddh2.2}
\end{eqnarray}
There are two ways to place two dd-vertices and two h-vertices on one 
suffix-loop. See Fig.9, where a small circle is used
to represent a dd-vertex explicitly.
\begin{figure}
\centerline{\epsfysize=4cm\epsfbox{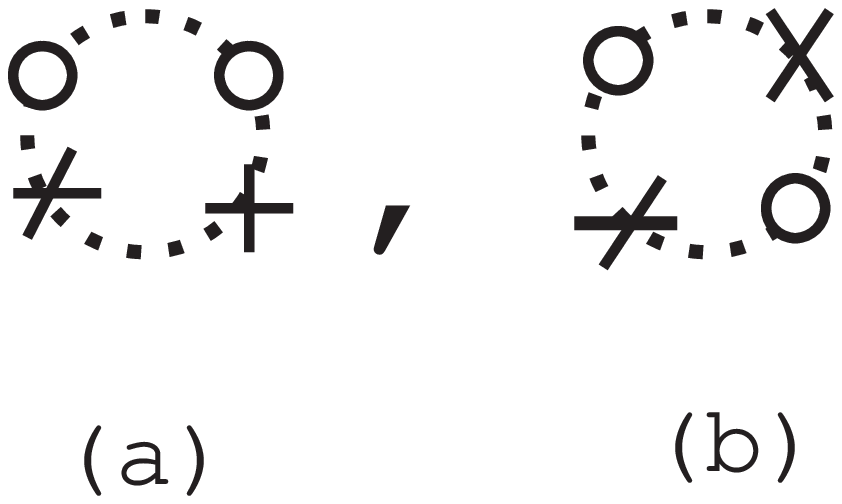}}
   \begin{center}
Fig.9\ 
Two ways to place two dd-vertices ( small circles) and two h-vertices
(cross marks) upon one suffix-loop.
   \end{center}
\end{figure}   
\begin{description}
\item[Def 4]\cite{II2}\ 
We call diagrams without bonds, like Fig.9, {\it bondless diagrams}. 
\end{description}
Finally,
taking account of the two bonds, we have three independent 
$(\pl\pl h)^2$-invariants for the case $l=1$. We name them $A1, A2$ and
$A3$ as shown in Fig.10.
\begin{figure}
\centerline{\epsfysize=4cm\epsfbox{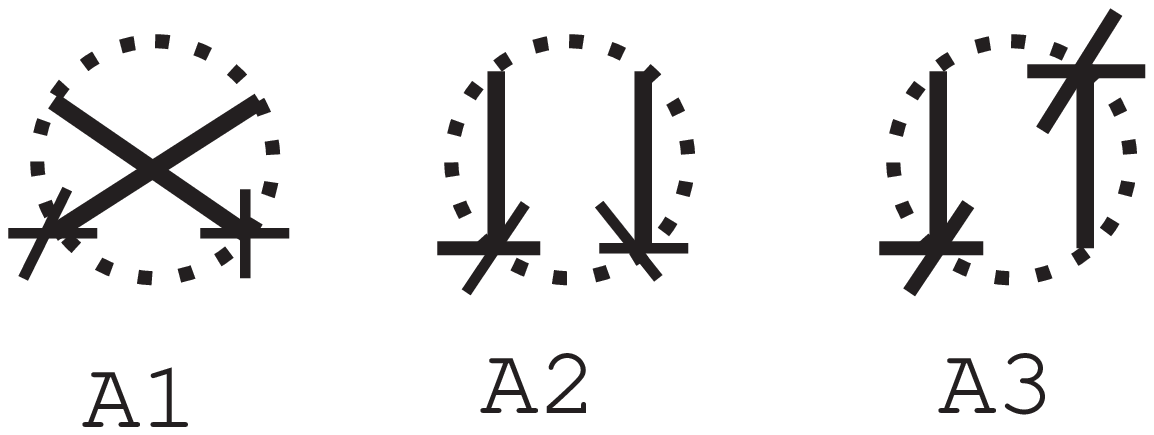}}
   \begin{center}
Fig.10\ 
Three independent $(\pl\pl h)^2$-invariants for the case of one suffix-loop.
   \end{center}
\end{figure}   
\flushleft{(ii) $l=2$}

\q For this case, we have 
\begin{eqnarray}
\left\{
\left(\begin{array}{c}  v_1 \\ w_1  \end{array}\right)\ 
\left(\begin{array}{c}  v_2 \\ w_2  \end{array}\right)
\right\}
=
(a):
\left(\begin{array}{c}  2 \\ 0  \end{array}\right)\ 
\left(\begin{array}{c}  0 \\ 2  \end{array}\right)\com\ 
(b):
\left(\begin{array}{c}  1 \\ 1  \end{array}\right)\ 
\left(\begin{array}{c}  1 \\ 1  \end{array}\right)\com\nn\\ 
(c):
\left(\begin{array}{c}  1 \\ 0  \end{array}\right)\ 
\left(\begin{array}{c}  1 \\ 2  \end{array}\right)\com\ 
(d):
\left(\begin{array}{c}  0 \\ 1  \end{array}\right)\ 
\left(\begin{array}{c}  2 \\ 1  \end{array}\right)\com\label{ddh2.3} 
\end{eqnarray}
where the order of
$\left(\begin{array}{c}  v_1 \\ w_1  \end{array}\right)\ $ and 
$\left(\begin{array}{c}  v_2 \\ w_2  \end{array}\right)\ $
is irrelevent for the present classification as stated in Lemma 2
\footnote{
The same treatment is adopted in the following other cases.
}
.
Each one above has one bondless diagram as shown in Fig.11.
\begin{figure}
\centerline{\epsfysize=7cm\epsfbox{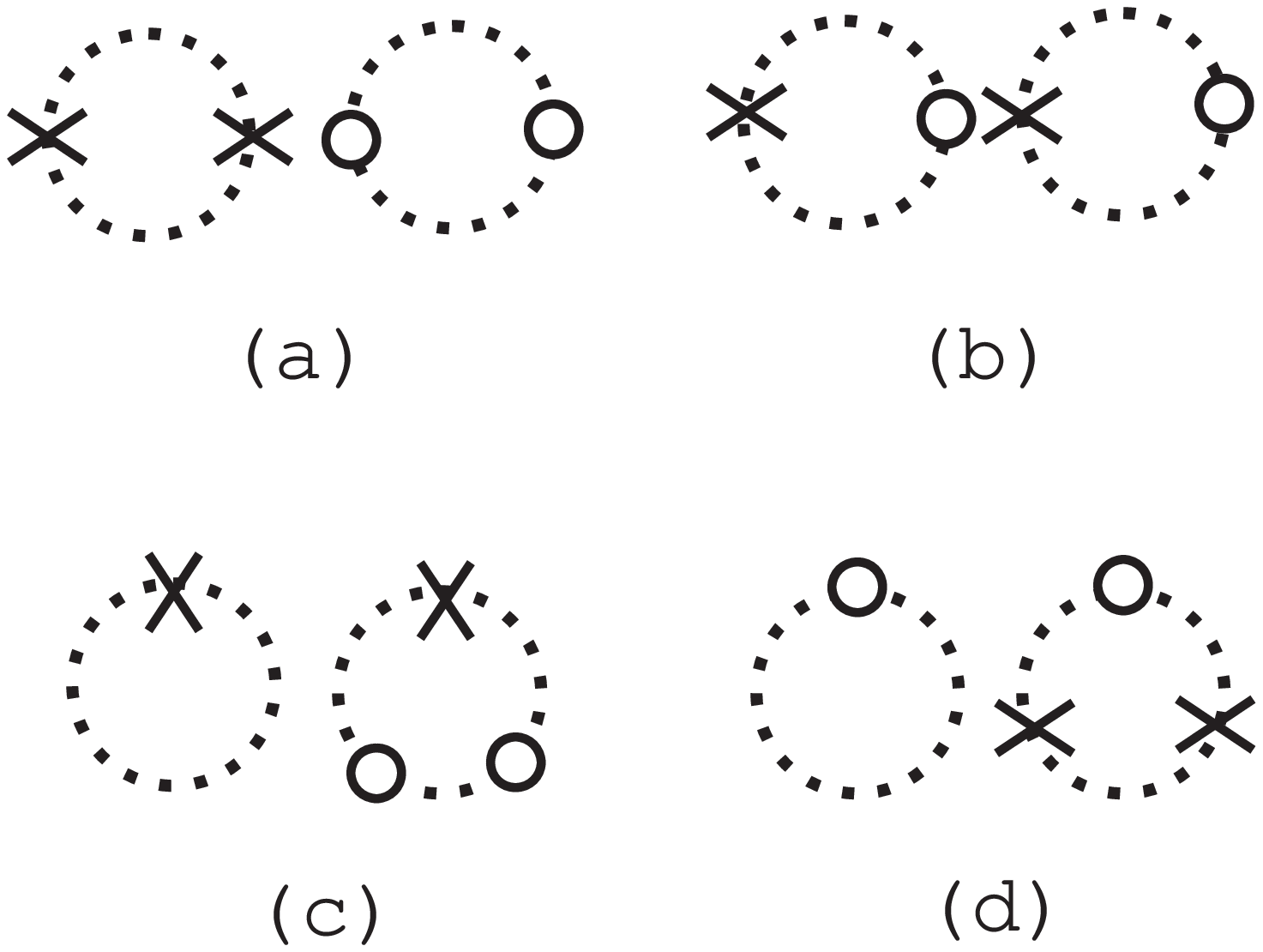}}
   \begin{center}
Fig.11\ 
Bondless diagrams for (\ref{ddh2.3}).
   \end{center}
\end{figure}
Then we have 5 independent $(\pl\pl h)^2$-invariants for this case
$l=2$. 
(Fig.11b has two independent ways to connect vertices by two bonds.) 
We name them $B1,\,B2,\,B3,\,B4$\, and $QQ$\, as shown in Fig.12.
Among them $QQ$\, is a {\it disconnected diagram}. 
\begin{figure}
\centerline{\epsfysize=7cm\epsfbox{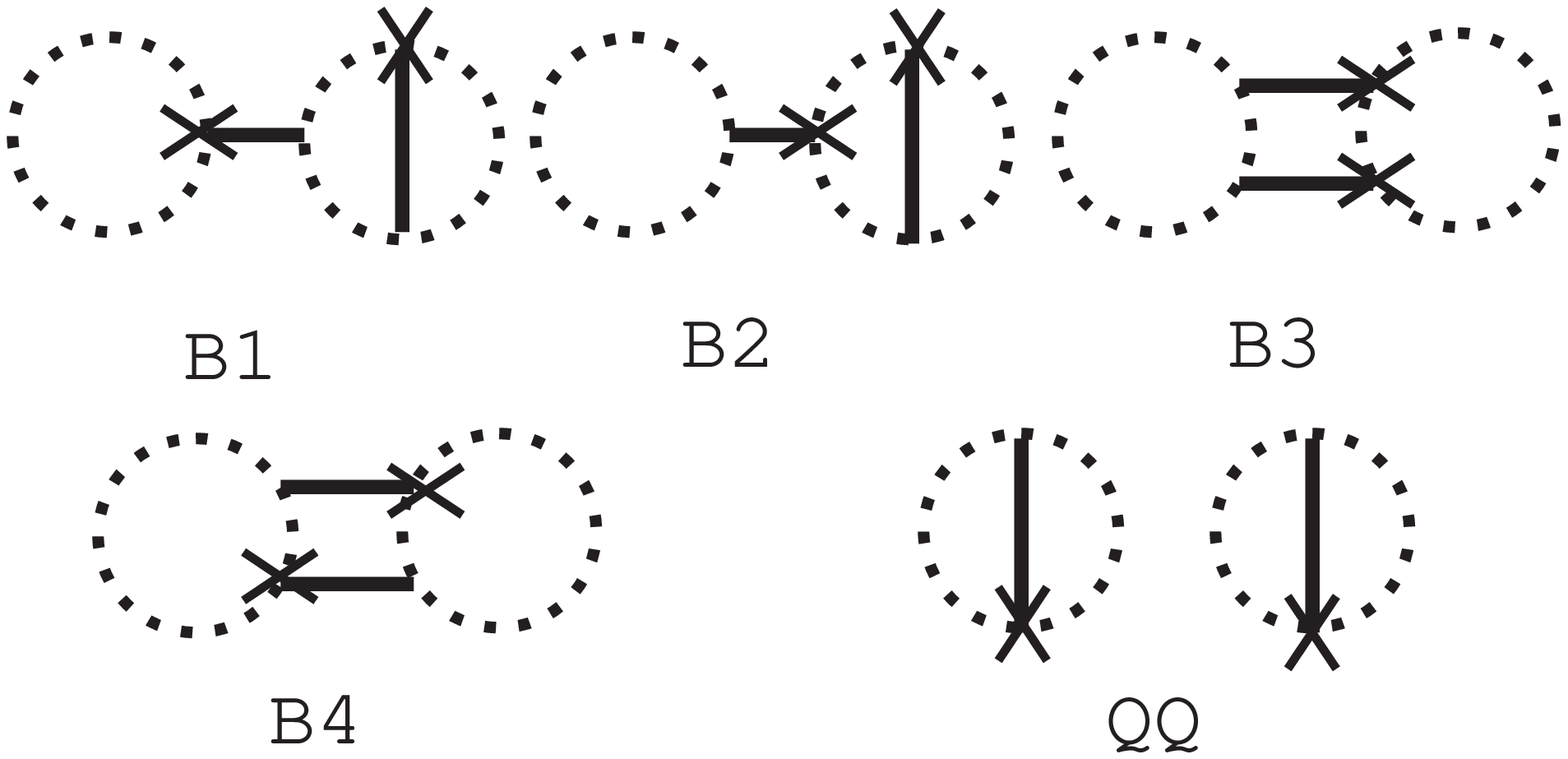}}
   \begin{center}
Fig.12\ 
Five independent $(\pl\pl h)^2$-invariants for the case of two suffix-loops.
   \end{center}
\end{figure}
\flushleft{(iii) $l=3$}

\q For this case, we have 
\begin{eqnarray}
\left\{
\left(\begin{array}{c}  v_1 \\ w_1  \end{array}\right)\ 
\left(\begin{array}{c}  v_2 \\ w_2  \end{array}\right)\ 
\left(\begin{array}{c}  v_3 \\ w_3  \end{array}\right)
\right\}
=
(a):
\left(\begin{array}{c}  1 \\ 0  \end{array}\right)\ 
\left(\begin{array}{c}  1 \\ 0  \end{array}\right)\ 
\left(\begin{array}{c}  0 \\ 2  \end{array}\right)\com\nn\\
(b):
\left(\begin{array}{c}  0 \\ 1  \end{array}\right)\ 
\left(\begin{array}{c}  0 \\ 1  \end{array}\right)\ 
\left(\begin{array}{c}  2 \\ 0  \end{array}\right)\com\ 
(c):
\left(\begin{array}{c}  1 \\ 0  \end{array}\right)\ 
\left(\begin{array}{c}  0 \\ 1  \end{array}\right)\ 
\left(\begin{array}{c}  1 \\ 1  \end{array}\right)\pr\label{ddh2.4} 
\end{eqnarray}
Each one above has one bondless diagram as shown in Fig.13.
\begin{figure}
\centerline{\epsfysize=7cm\epsfbox{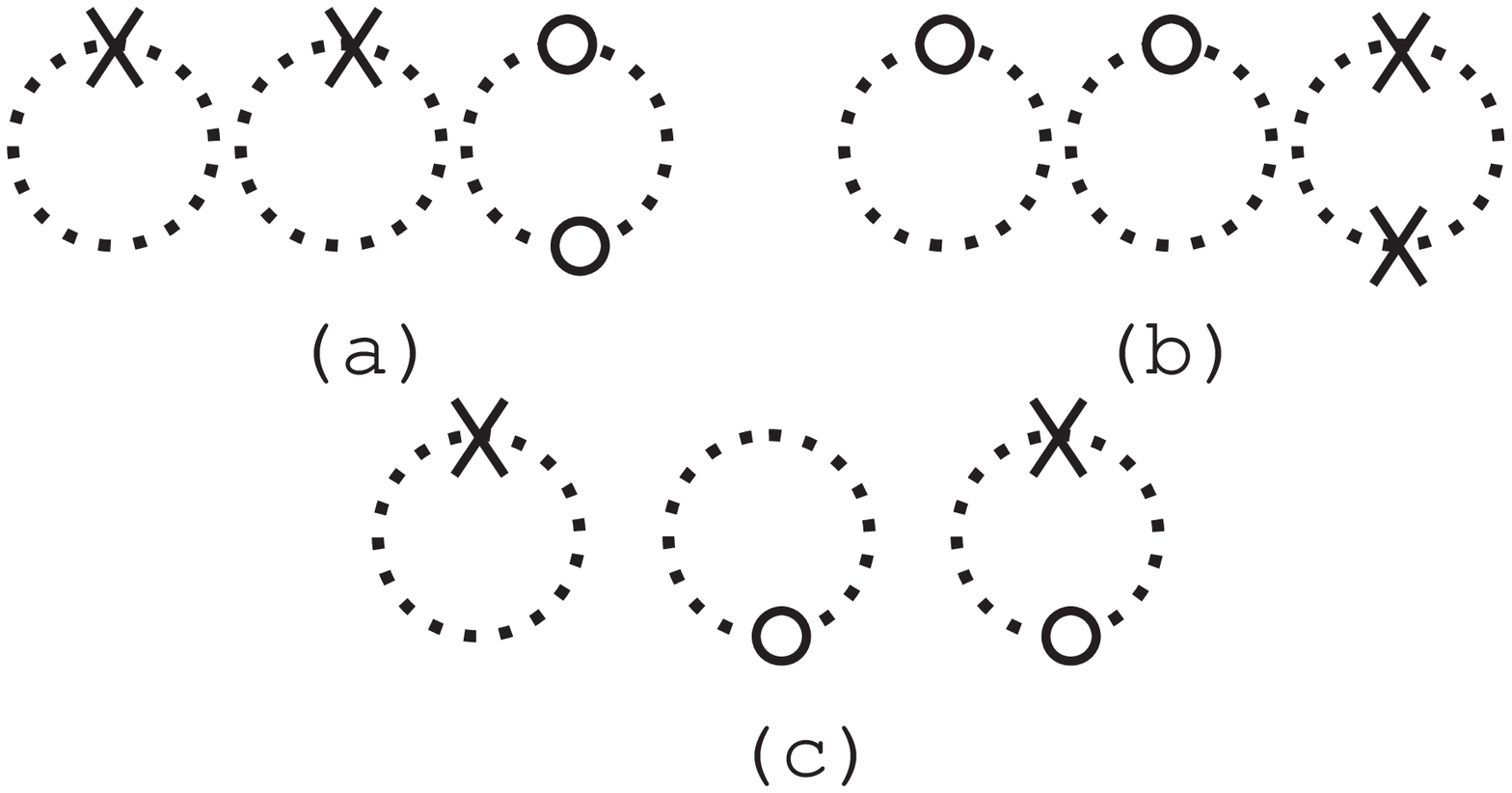}}
   \begin{center}
Fig.13\ 
Three bondless diagrams corresponding to (\ref{ddh2.4}).
   \end{center}
\end{figure}
Then we have 4 independent $(\pl\pl h)^2$-invariants for the case
$l=3$. 
(Fig.13c has two independent ways to connect vertices by two bonds.) 
We name them $C1,\,C2,\,C3,\,$\, and $PQ$\, as shown in Fig.14.
Among them $PQ$\, is a  disconnected diagram. 
\begin{figure}
\centerline{\epsfysize=7cm\epsfbox{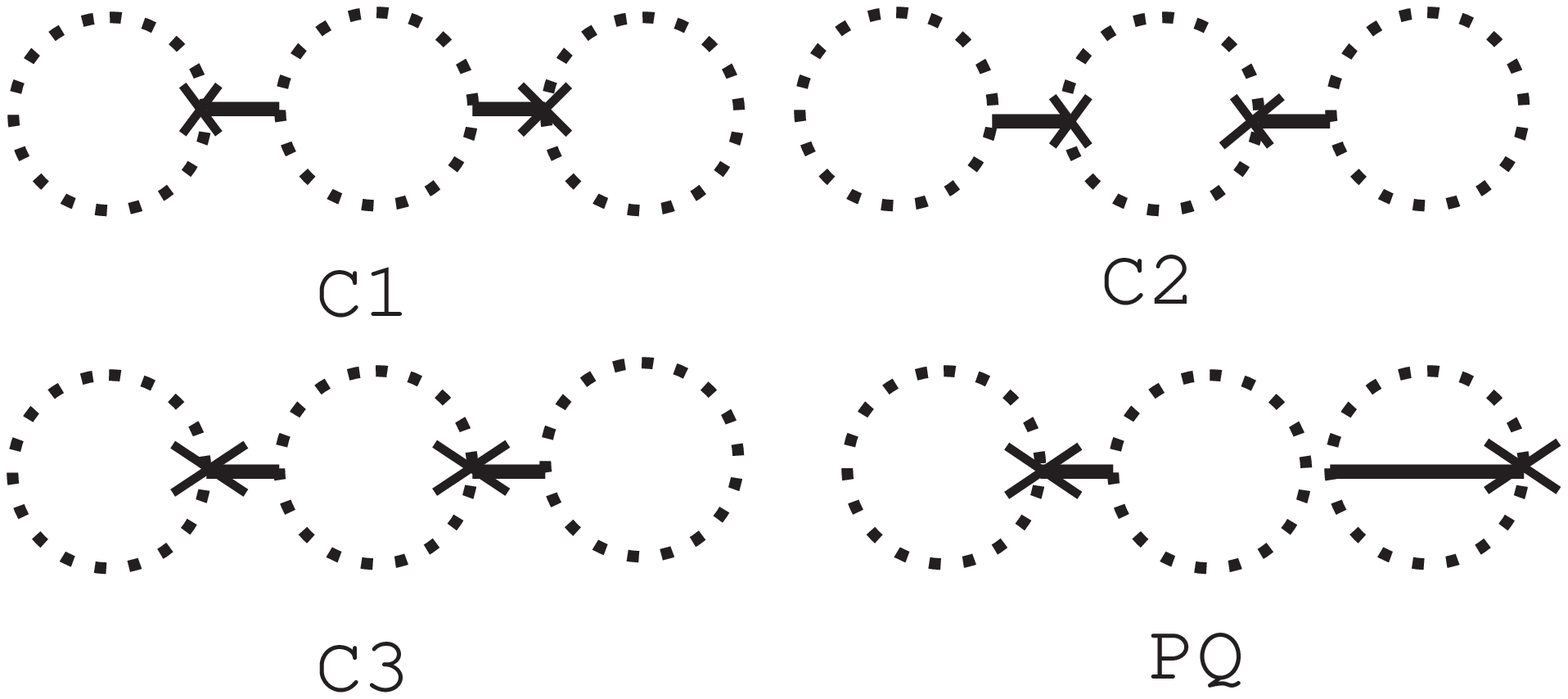}}
   \begin{center}
Fig.14\ 
Four independent $(\pl\pl h)^2$-invariants for the case of three suffix-loops.
   \end{center}
\end{figure}
\flushleft{(iv) $l=4$}

\q For this case, we have 
\begin{eqnarray}
\left\{
\left(\begin{array}{c}  v_1 \\ w_1  \end{array}\right)\ 
\left(\begin{array}{c}  v_2 \\ w_2  \end{array}\right)\ 
\left(\begin{array}{c}  v_3 \\ w_3  \end{array}\right)\ 
\left(\begin{array}{c}  v_4 \\ w_4  \end{array}\right)
\right\}
=
\left(\begin{array}{c}  1 \\ 0  \end{array}\right)\ 
\left(\begin{array}{c}  1 \\ 0  \end{array}\right)\ 
\left(\begin{array}{c}  0 \\ 1  \end{array}\right)\ 
\left(\begin{array}{c}  0 \\ 1  \end{array}\right)
\pr\label{ddh2.5} 
\end{eqnarray}
This corresponds to one bondless diagram  shown in Fig.15.
\begin{figure}
\centerline{\epsfysize=2cm\epsfbox{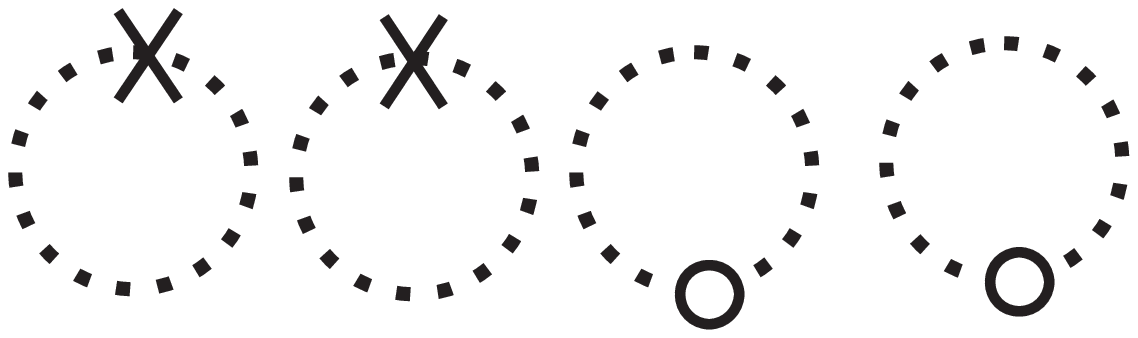}}
   \begin{center}
Fig.15\ 
The bondless diagram corresponding to (\ref{ddh2.5}).
   \end{center}
\end{figure}
Then we have a unique independent $(\pl\pl h)^2$-invariant (disconnected)
for the case
$l=4$. We name it $PP$\, as shown in Fig.16.
\begin{figure}
\centerline{\epsfysize=4cm\epsfbox{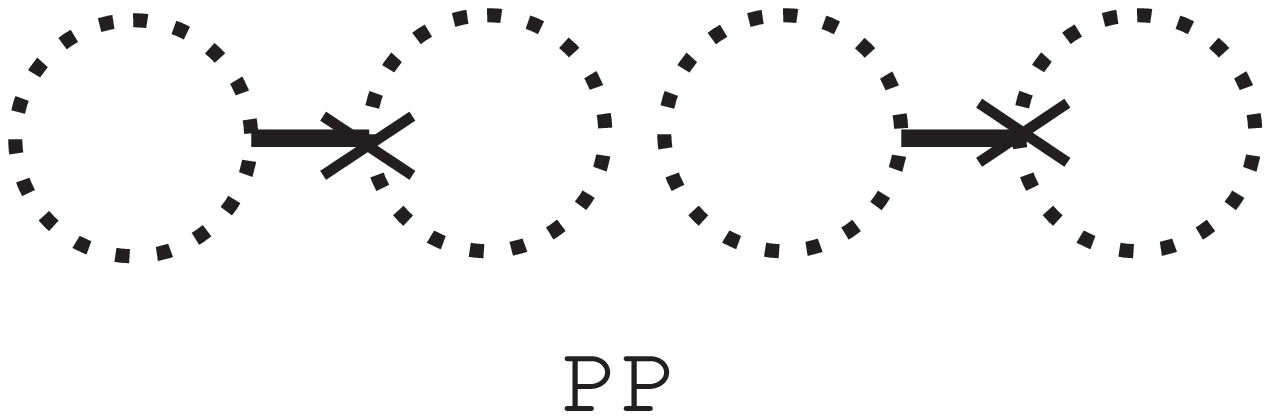}}
   \begin{center}
Fig.16\ 
The unique independent $(\pl\pl h)^2$-invariant 
for the case of four suffix-loops.
   \end{center}
\end{figure}

\q From (i)-(iv), we have obtained 
3($l=1$)+5($l=2$)+4($l=3$)+1($l=4$)=13 $(\pl\pl h)^2$-invariants using 
the {\it necessary}
conditions (\ref{ddh2.1}), Lemma 2.
( Among them, three ones (QQ,PQ,PP) are disconnected.) 
Their independence is assured by their difference of the 
connectivity of suffix-lines, in other words, of the {\it topology} of graphs. 
Thus we have completely listed up all independent $(\pl\pl h)^2$-invariants.
The ordinary mathematical expressions for the 13 invariants will be listed
in Table 1 of Sec.7.
In the next section, we
reprove the completeness of the above enumeration from the standpoint of
a suffix-permutation symmetry and the combinatorics among suffixes.
In the similar way we can list up all independent invariants
for the general type :\ $(\pl\pl h)^s$.

\q Let us summarize this section by giving the algorithm to compute
all  $(\pl\pl h)^s$-invariants.
\begin{description}
\item[Step G1]\ For each $l(=1,2,\cdots,2s)$, obtain all possible
sets of 
$\left\{
\left(\begin{array}{c}  v_i \\ w_i  \end{array}\right)\ ;\ 
i=1,2,\cdots,l-1,l
\right\}$
from the {\it necessary} condition (\ref{ddh2.1}).
\item[Step G2]\ For each $l$ and each set 
$\left\{
\left(\begin{array}{c}  v_i \\ w_i  \end{array}\right)\ ;\ 
i=1,2,\cdots,l-1,l
\right\}$
, list all independent bondless diagrams.
\item[Step G3]\ For each bondless diagram, list all independent 
graphs obtained by "placing" bonds on it in all possible ways.
\end{description}

\section{Completeness of Graph Enumeration }
\q Let us examine the $\pl\pl h$- and $(\pl\pl h)^2$-invariants 
from the viewpoint of the suffix-permutation symmetry 
stated below (\ref{Rep.3}).

\flushleft{(i) $\pl\pl h$-invariants}

\q The $\pl\pl h$-invariants are obtained by contracting 4 indices
$(\m_1,\m_2,\m_3,\m_4)$ in $\pl_{\m_1}\pl_{\m_2}h_{\m_3\m_4}$.
All possible ways of contracting the four indices 
are given by the following 3 ones.
\begin{eqnarray}
a)\ \del_{\m_1\m_2}\del_{\m_3\m_4}\com\q
b)\ \del_{\m_1\m_3}\del_{\m_2\m_4}\com\q
c)\ \del_{\m_1\m_4}\del_{\m_2\m_3}
\pr\label{gc.1}
\end{eqnarray}
Due to the symmetry given below (\ref{Rep.3}), 
we see b) and c) give the same invariant Q.
\begin{description}
\item[Def 5]\cite{II2}\ 
We generally call the number of occurrence of a covariant
(which includes the case of an invariant) C, when contracting
suffixes of a covariant C$'$ in all possible ways, a {\it weight} of C
from C$'$.
\end{description}
In the present case, P has a weight 1 and Q has a weight 2 (from 4-tensor
$\pl_\m\pl_\n h_\ab$).
We have an identity
between the number of all possible ways of suffix-contraction 
(\ref{gc.1}) and weights
of invariants.
\begin{eqnarray}
3=1(P)+2(Q)
\pr\label{gc.2}
\end{eqnarray}
A weight of an invariant shows  'degeneracy' in the contraction due to
its suffix-permutation symmetry.
The above identity shows 
the completeness of the enumeration of $\pl\pl h$-invariants from the viewpoint
of the permutation symmetry.

\flushleft{(ii) $(\pl\pl h)^2$-invariants}

\q We can do the same analysis for $(\pl\pl h)^2$-invariants. The number of all
possible contraction of 8 indices in the 8-tensor 
$\pl_{\m_1}\pl_{\m_2}h_{\m_3\m_4}\cdot\pl_{\m_5}\pl_{\m_6}h_{\m_7\m_8}$
is \  $7\times 5\times 3\times 1=105$
, from (\ref{sumweight}) with $N=8$. 
Let us take $B1$ of Fig.12 as an example
of weight calculation. See Fig. 17.
\begin{eqnarray}
\mbox{The weight of B1 from the 8-tensor}
 = 1\mbox{(weight of Fig.2b from 4-tensor $\pl\pl h$)}\nn\\
\times 4\mbox{(weight of Fig.2c from 4-tensor $\pl\pl h$)}
\times 2\mbox{(two ways of 2b-2c contraction)}       \nn\\
\times 2\mbox{(two ways of choosing 2b-bond and 2c-bond among 2 bonds)}=16
\pr\label{gc.3}
\end{eqnarray}

\begin{figure}
\centerline{\epsfysize=4cm\epsfbox{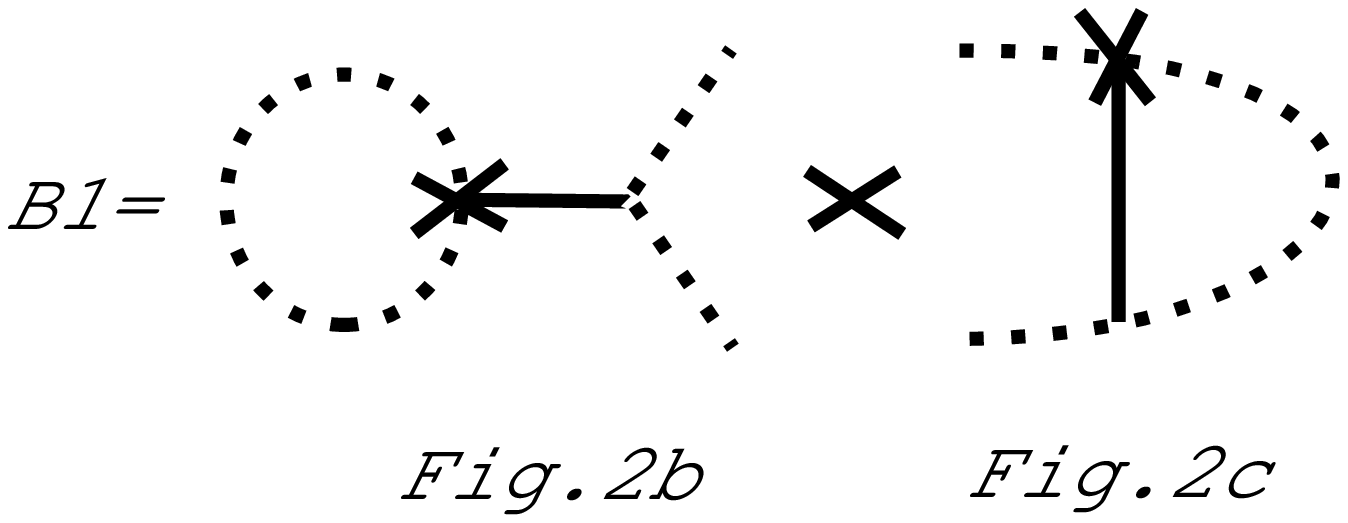}}
   \begin{center}
Fig.17\ 
Graph B1 for the weight calculation (\ref{gc.3}).
   \end{center}
\end{figure}

Similarly we can obtain  weights for all other $(\pl\pl h)^2$-invariants
and the following identity holds true.
\begin{eqnarray}
105=7\times 5\times 3\times 1=16(A1)+16(A2)+16(A3)\nn\\
+16(B1)+16(B2)+4(B3)+4(B4)+4(QQ)\q\label{gc.4}\\
+2(C1)+2(C2)+4(C3)+4(PQ)+1(PP)\pr\nn
\end{eqnarray}
This identity clearly 
shows the completeness of the 13 $(\pl\pl h)^2$-invariants 
listed in Sect.4 and 5.

\q Weights, defined above, correspond to the symmetry factor or the statistical
factor in the Feynman diagram expansion of the field theory (Sec.3). Further the
above identity (\ref{gc.4}) reminds us of a similar one, in the graph theory,
called 'Polya's enumeration theorem\cite{H}.

\section{Topological Indices of Graphs} 
\q The graph representation is very useful in proving 
mathematical properties, such as completeness
and independence, of SO(n) invariants because the connectivity of
suffixes can be read in the topology of a graph. In practical calculation,
however, depicting graphs is cumbersome.
In order to specify
every graph of invariant succinctly, we introduce a set of 
{\it topological indices} which
show  how  suffix-lines (suffixes) 
are connected (contracted) in a graph (an invariant). 
In this section we characterize every independent graph of invariant
of types $\pl\pl h$ and $(\pl\pl h)^2$,  
by a set of some topological indices
\footnote{
This approach is very contrasting, in the sense that they also
express a graph in terms of a set of numbers, 
with standard ones in the graph theory:\ 
the incidence matrix or the adjacency matrix (Sec.4) 
\cite{H,N}. In Sec.8 we will present a relation
to the adjacency matrix.
}
. 

\flushleft{(i)\ Number of Suffix Loops (\ul{$l$})}\footnote{
All indices are underlined. For example, \ul{$l$}, \ul{tadpoleno},
\ul{tadtype}, \ul{bcn}, and \ul{vcn}.
}

\q The number of suffix loops (\ul{$l$}) of a graph is a good index. In fact, 
every $\pl\pl h$-invariant is completely characterized by \ul{$l$}:\ \ 
\ul{$l$}=2 for P and \ul{$l$}=1 for Q. The index \ul{$l$} is not sufficient
to discriminate every $(\pl\pl h)^2$-invariant. We need the following ones, 
(ii) and (iii).

\flushleft{(ii)\ Number of Tadpoles (\ul{tadpoleno}) and Type of 
Tadpole (\ul{tadtype}[\ ])}

\begin{description}
\item[Def 6]\cite{II2}\ 
We call a closed suffix-loop which has only one vertex, a {\it tadpole}. 
The number of tadpoles which a graph has, is called {\it tadpole number}
(\ul{tadpoleno}) of the graph. Let us consider $t$-th tadpole in a graph 
($t=1,2,\cdots,$\ul{tadpoleno}). 
When the vertex which the $t$-th tadpole has, is dd-vertex (h-vertex)
its {\it tadpole type}, \ul{tadtype}[$t$], is defined
to be 0 (1). \ul{tadtype}[\ ] is assigned for each tadpole.
\end{description}

For example, 
in Fig.3, P has \ul{tadpoleno}=2 and \ul{tadtype}[1]=0 and 
\ul{tadtype}[2]=1.
\footnote{
Final results should be independent of arbitrariness in numbering all
tadpoles in a graph.
}
Q has \ul{tadpoleno}=0.

\flushleft{(iii)\ Bond Changing Number(\ul{bcn}[\ ]) and 
Vertex Changing Number(\ul{vcn}[\ ])}

\begin{description}
\item[Def 7]\cite{II2}\ 
\ul{bcn}[$k$] and \ul{vcn}[$k$] are defined for $k$-th suffix-loop 
($k=1,2,\cdots,$\ul{$l$}) as follows.
When we trace a suffix-loop, starting from a vertex 
in a certain direction, we generally pass some vertices,
and  finally come back to the starting vertex. See Fig.18.
When we move, in the tracing, from a vertex to a next vertex, 
we compare the bonds to which the two vertices belong, and their vertex-types.
If the bonds are different, we set $\Del\mbox{\ul{bcn}}=1$, otherwise
$\Del\mbox{\ul{bcn}}=0$, 
If the vertex-types are different, we set $\Del\mbox{\ul{vcn}}=1$, otherwise
$\Del\mbox{\ul{vcn}}=0$. 
For $k$-th loop, we sum every number of $\Del\mbox{\ul{bcn}}$ and
$\Del\mbox{\ul{vcn}}$ while tracing the loop once, and
assign as
$\sum_{\mbox{along $k$-th loop}}\Delta \mbox{\ul{bcn}}\equiv$ \ul{bcn}[$k$],
$\sum_{\mbox{along $k$-th loop}}\Delta \mbox{\ul{vcn}}\equiv$ \ul{vcn}[$k$].
\footnote{
Final results should be independent of arbitrariness of numbering all
suffix-loops in a graph.
}
\end{description}
A practical way to calculate \ul{bcn}[\ ] and \ul{vcn}[\ ] is
explained in Appendix C.

\q In Table 1, we list all indices necessary
for discriminating every $(\pl\pl h)^2$-invariant completely. 
The listed 13 invariants are independent each other because Table 1
clearly shows the topology of every graph is different.

\newpage
\begin{tabular}{|c||c|c|c|c|c|}
\hline
Graph\ $\backslash$\ Indices
& \ul{$l$}& \ul{tadpoleno}& \ul{tadtype}[\ ]&\ul{bcn}[\ ]&\ul{vcn}[\ ]\\
\hline
        &       &            &              &              &              \\
$A1
=\pl_\si\pl_\la h_\mn\cdot\pl_\si\pl_\n h_{\m\la}$ 
        &  1    & 0          & nothing      & 4            & 2            \\
        &       &            &              &              &              \\
\hline
        &       &            &              &              &              \\
$A2
=\pl_\si\pl_\la h_{\la\m}\cdot\pl_\si\pl_\n h_{\mn}$
        & 1    & 0          & nothing      & 2            & 2            \\
        &       &            &              &              &              \\
\hline
        &       &            &              &              &              \\
$A3
=\pl_\si\pl_\la h_{\la\m}\cdot\pl_\m\pl_\n h_{\n\si}$
        & 1    & 0          & nothing      & 2            & 4            \\
        &       &            &              &              &              \\
\hline
\hline
        &       &            &              &              &              \\
$B1
=\pl_\n\pl_\la h_{\si\si}\cdot\pl_\la\pl_\m h_{\mn}$
        & 2    & 1          & 1            & $/$          & $/$          \\
        &       &            &              &              &              \\
\hline
        &       &            &              &              &              \\
$B2
=\pl^2 h_{\la\n}\cdot\pl_\la\pl_\m h_{\mn}$
        & 2    & 1          & 0            & $/$          & $/$          \\
        &       &            &              &              &              \\
\hline
        &       &            &              &  2          &  0          \\
\cline{5-6}
$B3
=\pl_\m\pl_\n h_{\la\si}\cdot\pl_\m\pl_\n h_{\ls}$
        & 2    & 0          & nothing      &   2          &  0          \\
\cline{5-6}
        &       &            &              &              &              \\
\hline
        &       &            &              &   2         &  2          \\
\cline{5-6}
$B4
=\pl_\m\pl_\n h_{\la\si}\cdot\pl_\la\pl_\si h_{\mn}$
        & 2    & 0          & nothing      &    2         &  2          \\
\cline{5-6}
        &       &            &              &              &              \\
\hline
        &       &            &              &   0         &  2           \\
\cline{5-6}
$Q^2
=(\pl_\m\pl_\n h_{\mn})^2$
        & 2     & 0          & nothing      &   0         &  2           \\
\cline{5-6}
        &       &            &              &              &               \\
\hline
\hline
        &       &            & 1             &              &                \\
$C1
=\pl_\m\pl_\n h_{\la\la}\cdot\pl_\m\pl_\n h_{\si\si}$
        & 3    & 2          & 1          &   $/$         &  $/$           \\
        &       &            &              &              &                \\
\hline
        &       &            &  0            &              &                \\
$C2
=\pl^2 h_{\mn}\cdot\pl^2 h_{\mn}$
        & 3    & 2          &   0          &   $/$         &  $/$           \\
        &       &            &              &              &                \\
\hline
        &       &            &  1            &   0         &  0           \\
\cline{4-6}
$C3
=\pl_\m\pl_\n h_{\la\la}\cdot\pl^2 h_{\mn}$
        & 3   & 2          & 0          &   0         &  0          \\
\cline{4-6}
        &       &            &              &   2         &  2          \\
\hline
        &       &            &     1         &   0         &  0           \\
\cline{4-6}
$PQ
=\pl^2 h_{\la\la}\cdot\pl_\m\pl_\n h_{\mn}$
        & 3   & 2          & 0          &   0         &  0          \\
\cline{4-6}
        &       &            &              &   0         &  2           \\
\hline
\hline
        &       &            &              &              &              \\
$P^2
=(\pl^2 h_{\la\la})^2$
        & 4  &  $/$       & $/$          & $/$          & $/$          \\
        &       &            &              &              &              \\
\hline
\multicolumn{6}{c}{\q}                                                 \\
\multicolumn{6}{c}{Table 1\ \  List of indices for all 
$(\pl\pl h)^2$-invariants. 
The symbol '$/$' means }\\
\multicolumn{6}{c}{'need not be calculated for discrimination'. 
}\\
\end{tabular}

\vs 1
\begin{figure}
\centerline{\epsfysize=8cm\epsfbox{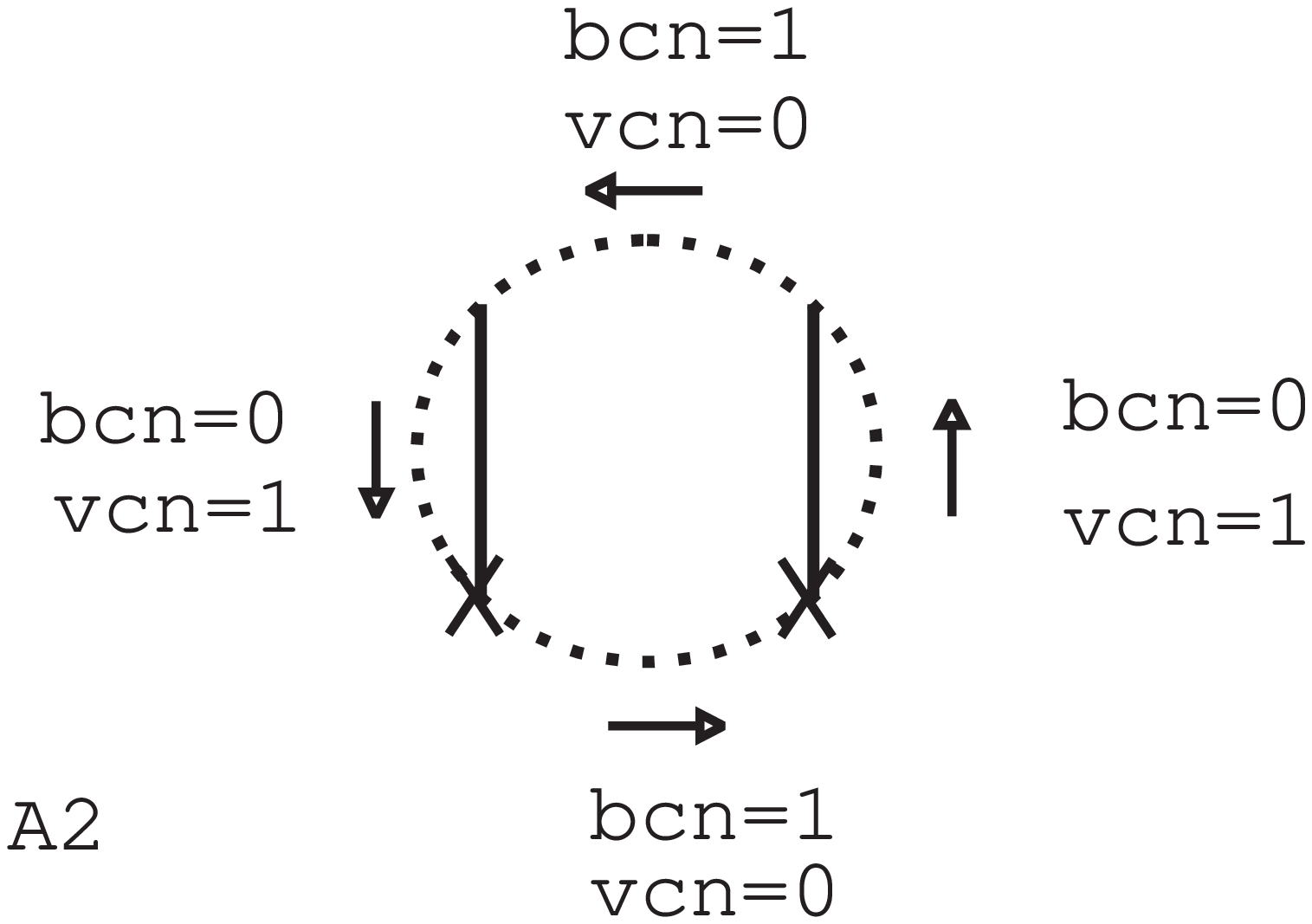}}
   \begin{center}
Fig.18\ Explanation of \ul{bcn}[\ ] and \ul{vcn}[\ ] using Graph A2.
   \end{center}
\end{figure}

\q We have presented the set of indices 
which is sufficient to discriminate
all $\pl\pl h$- and $(\pl\pl h)^2$-invariants. For the case 
of $(\pl\pl h)^3$-invariants, see Ref.\cite{II2}. 
It is a non-trivial work to find an appropriate set of indices
for higher-order invariants. However, once the set of indices is fixed, 
the computation itself is very efficient and is most appropriate
for the computer algorithm\cite{SI96}. These advatageous 
and disadvatageous points should be compared with the case of
the adjacency matrix. In Sec.8, we will explain how to read
the indices from the adjacency matrix.

\section{Calculation of Indices from Adjacency Matrices}
In Sec.4 we have introduced the adjacency matrix to represent
a graph. As shown in Step M3, there exist some equivalent matrices
which express the same graph due to the arbitrariness of vertex-naming.
It is very hard (at least practically) to find the permutation $P$ in
(\ref{ad.4}) and identify a representative.
In this section, we explain how to read the topological indices of
a graph from the matrix:\ 
$A=[a_{IJ}];\ I,J=1,{\bar 1},2,{\bar 2},\cdots,B,{\bar B}$. 
It is useful when 
we efficiently identify an invariant ( a representative )
in the matrix representation. 

\flushleft{(i)\ Number of Tadpoles (\ul{tadpoleno}) and Type of 
Tadpole (\ul{tadtype}[\ ])}

\q These two indices can be most easily read from the matrix.

\begin{eqnarray}
\mbox{\ul{tadpoleno}}=\half\sum_{I} a_{II}=\half~\mbox{Trace}~A\ ,\ 
                         I=1,{\bar 1},2,{\bar 2},\cdots,B,{\bar B}       \pr\nn\\
\mbox{\ul{tadtype}[t]}=\left\{ \begin{array}{ll}
         0\ \mbox{(der-vertex)}\q & \mbox{for } I=i\ ,\ a_{II}\neq 0\\
         1\ \mbox{(h-vertex)}\q & \mbox{for } I={\bar i}\ ,\ a_{II}\neq 0
                \end{array}
		\right.		                   \com\nn\\
		t=1,2,\cdots,\mbox{\ul{tadpoleno}}\pr
\label{matind.1}
\end{eqnarray}
As example, for $C3$ of (\ref{Mddhddh}) we have
\begin{eqnarray}
\mbox{\ul{tadpoleno}}=\half (2+0+0+2)=2            \com\nn\\
\mbox{\ul{tadtype}[1]}=0\ (\ a_{11}=2\ )           \com\nn\\      
\mbox{\ul{tadtype}[2]}=1\ (\ a_{{\bar 2}{\bar 2}}=2\ )           \com      
\label{matind.2}
\end{eqnarray}
which are same as shown in Table 1.

\vs{0.5}
\flushleft{(ii)\ \ul{connectivity}}

First we define a new index {\it connectivity}.
\begin{description}
\item[Def 8]\cite{II2}\ 
Let us consider a graph of invariants with $s$ bonds. 
There are $\mbox{}_sC_2=s(s-1)/2$ different pairs of bonds. We define
\ul{connectivity} of the graph as  
the total number of those pairs which are connected by at least
one suffix-line. 
$0\leq \mbox{\ul{connectivity}} \leq s(s-1)/2$. 
\end{description}

This new index is one of important indices in the calculation of
the $(\pl\pl h)^3$-invariants \cite{II2}. An adjacency matrix
with the size of $2B\times 2B$ is composed of $B^2$ submatrices
$B_{ij}$ with the size of $2\times 2$. Then the connectivity is
obtained by
\begin{eqnarray}
& \mbox{\ul{connectivity}}=\half \{ \ \mbox{No. of off-diagonal}(i\neq j)\ 
\mbox{non-empty elements }B_{ij}\ \mbox{in}\ A=(B_{ij})\ \}
            \ ,& \nn\\
& B_{ij}\equiv 
\left[
\matrix{
a_{ij}        & a_{i{\bar j}}        \cr
a_{{\bar i}j} & a_{{\bar i}{\bar j}} \cr
}
\right]=(B_{ji})^T\com &
\label{matind.3}
\end{eqnarray}
where "empty" ("non-empty") means 
${a_{ij}}^2+{a_{i{\bar j}}}^2+{a_{{\bar i}j}}^2+{a_{{\bar i}{\bar j}}}^2
=(\neq) 0$. When $B_{ij}$ is empty, it means $i$-th bond and $j$-th bond
are not connected by any suffix-lines. We give an example.
\begin{eqnarray}
G9\equiv \pl_\m\pl_\om h_\mn\cdot\pl_\n\pl_\la h_{\tau\si}\cdot
\pl_\la\pl_\si h_{\tau\om}=\mbox{Fig.19}=
\left[
\matrix{
0 & 0 & 1 & 1 & 0 & 0 \cr
0 & 0 & 0 & 1 & 1 & 0 \cr
1 & 0 & 0 & 0 & 0 & 1 \cr
1 & 1 & 0 & 0 & 0 & 0 \cr
0 & 1 & 0 & 0 & 0 & 1 \cr
0 & 0 & 1 & 0 & 1 & 0 \cr
}
\right]\com\nn\\
B_{12}\ ,\ B_{23}\ ,\ B_{31}\ ,\ 
B_{21}={B_{12}}^T\ ,\ B_{32}={B_{23}}^T\ \mbox{and}\ B_{13}={B_{31}}^T
\ \mbox{are non-empty}\com \nn\\
\mbox{\ul{connectivity}}=\half\times 6=3\pr\qqq\qqq\qqq
\label{matind.4}
\end{eqnarray}
\begin{figure}
\centerline{\epsfysize=4cm\epsfbox{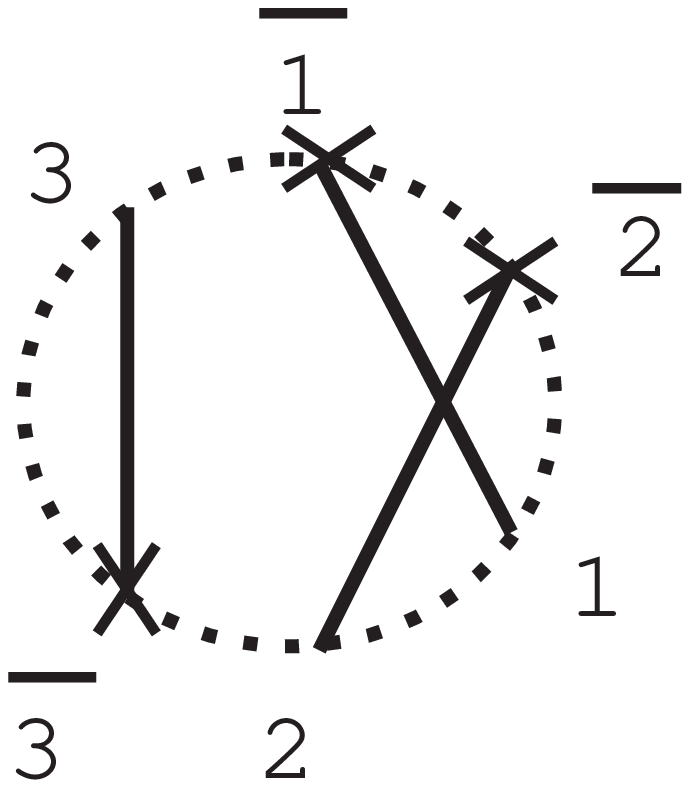}}
   \begin{center}
Fig.19\ Graph G9$\equiv \pl_\m\pl_\om h_\mn\cdot\pl_\n\pl_\la h_{\tau\si}
\cdot\pl_\la\pl_\si h_{\tau\om}$.
   \end{center}
\end{figure}
\ul{disconnectivity} is defined in Ref.\cite{II2} and can be
calculated from adjacency matrices in the similar way.

\vs{0.5}
\flushleft{(iii)\ Number of Suffix Loops (\ul{$l$})}\nl

\q There are several important indices associated with
a suffix-loop. Therefore it is the first important thing to find
every independent suffix-loop in an adjacency matrix.
Let us explain it with an example.
\begin{eqnarray}
G25\equiv \pl_\m\pl_\n h_\ls\cdot\pl_\si\pl_\tau h_\mn\cdot
\pl_\la\pl_\om h_{\tau\om}=\mbox{Fig.20}=
\left[
\matrix{
0 & 1_a & 0 & 1_{d'} & 0 & 0 \cr
1_{a'} & 0 & 0 & 0 & 1_b & 0 \cr
0 & 0 & 0 & 0 & 0 & 2_{e,f'} \cr
1_d & 0 & 0 & 0 & 1_{c'} & 0 \cr
0 & 1_{b'} & 0 & 1_c & 0 & 0 \cr
0 & 0 & 2_{f,e'} & 0 & 0 & 0 \cr
}
\right]\pr 
\label{matind.5}
\end{eqnarray}
\begin{figure}
\centerline{\epsfysize=8cm\epsfbox{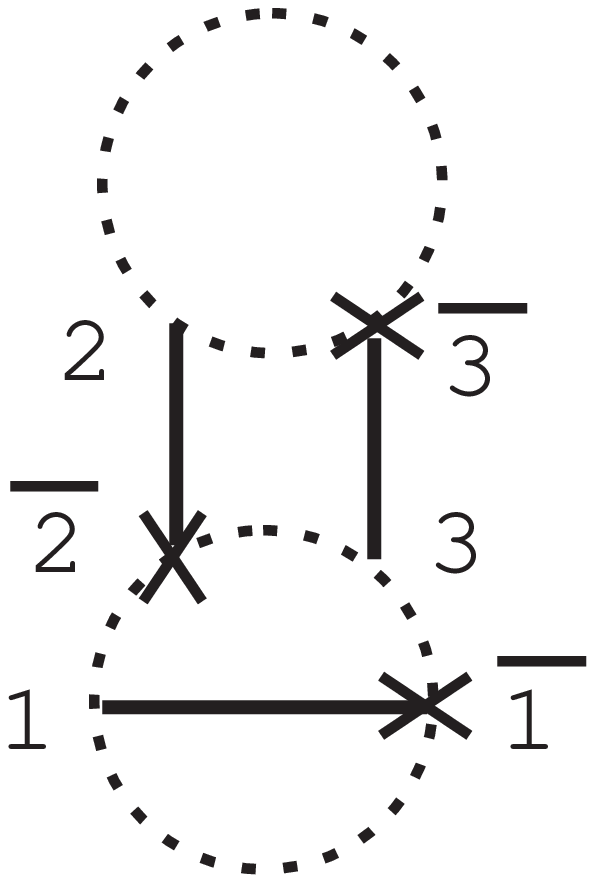}}
   \begin{center}
Fig.20\ Graph of G25$\equiv \pl_\m\pl_\n h_\ls\cdot\pl_\si\pl_\tau h_\mn
\cdot\pl_\la\pl_\om h_{\tau\om}$.
   \end{center}
\end{figure}

We see two independent suffix-loops,\ul{$l$}=2 :\ 
[loop1]\ $a-b-c-d-a$\ ;\ [loop2]\ $e-f-e$. 
The remaining two loops, $a'-b'-c'-d'-a'$\ 
and $e'-f'-e'$, are the symmetric copies of loop1 and loop2
respectively. ( loop2 and its copy overlap in the
matrix.) 
To systematically find suffix-loops 
in a finite-size matrix in general
is easy ( by use of a computer ) as far as the size is not so large.
It is done by searching closed loops of "successive" 
(row,column)-numbers of the adjacency matrix.
\vs {0.3}
\setlength{\unitlength}{1mm}
\begin{picture}(160,60)
\put(0,54){$\Del$\ul{bcn}}
\put(30,54){0}
\put(60,54){1}
\put(90,54){1}
\put(120,54){1}
\put(125,56){\vector(1,0){5}}
\put(135,54){3=\ul{bcn}[1]}
\put(0,48){$\Del$\ul{vcn}}
\put(30,48){1}
\put(60,48){1}
\put(90,48){1}
\put(120,48){1}
\put(125,50){\vector(1,0){5}}
\put(135,48){4=\ul{vcn}[1]}
\put(0,40){loop1}
\put(23,40){$a:(1,{\bar 1})$}
\put(40,42){\line(1,0){10}}
\put(53,40){$b:({\bar 1},3)$}
\put(70,42){\line(1,0){10}}
\put(83,40){$c:(3,{\bar 2})$}
\put(100,42){\line(1,0){10}}
\put(113,40){$d:({\bar 2},1)$}
\put(120,39){\line(0,-1){4}}
\put(120,35){\line(-1,0){90}}
\put(30,35){\line(0,1){4}}
\put(0,24){$\Del$\ul{bcn}}
\put(30,24){1}
\put(60,24){1}
\put(125,26){\vector(1,0){5}}
\put(135,24){2=\ul{bcn}[2]}
\put(0,18){$\Del$\ul{vcn}}
\put(30,18){1}
\put(60,18){1}
\put(125,20){\vector(1,0){5}}
\put(135,18){2=\ul{vcn}[2]}
\put(0,10){loop2}
\put(23,10){$e:(2,{\bar 3})$}
\put(40,12){\line(1,0){10}}
\put(53,10){$f:({\bar 3},2)$}
\put(60,9){\line(0,-1){4}}
\put(60,5){\line(-1,0){30}}
\put(30,5){\line(0,1){4}}
\end{picture}
\vs {0.3}
The numbers associated with $\Del$\ul{bcn} and $\Del$\ul{vcn}, in 
the above illustration, is used in the next item. 

\vs{0.5}
\flushleft{(iv)\ Bond Changing Number(\ul{bcn}[\ ]) and 
Vertex Changing Number(\ul{vcn}[\ ])}\nl

\q When a suffix-loop, which is named $l$-th loop, 
is given in the form of a series of
(row, column)-numbers of an adjacency matrix, as given in iii), 
$\Del$\ul{bcn} and $\Del$\ul{vcn} are immediately given by
as follows.
\begin{eqnarray}
\mbox{For an element (row=$I$,column=$J$)}\ \mbox{in the $l$-th loop}\nn\\
\Del\mbox{\ul{bcn}}=\left\{ \begin{array}{ll}
         0\  & \mbox{for } i=j\\
         1\  & \mbox{for } i\neq j
                             \end{array}
		              \right.		                   \com\nn\\
\Del\mbox{\ul{vcn}}=\left\{ \begin{array}{ll}
         0\  & \mbox{for } (I=i,J=j)\mbox{ or }(I={\bar i},J={\bar j})\\
         1\  & \mbox{for } (I=i,J={\bar j})\mbox{ or }(I={\bar i},J=j)
                             \end{array}
		              \right.		                   \pr
\label{matind.6}
\end{eqnarray}
The total sums of $\Del$\ul{bcn} and $\Del$\ul{vcn} 
along the $l$-th loop give
\ul{bcn}[$l$] and \ul{vcn}[$l$] respectively. An example is
given in iii).

\vs 1

\q Another interesting index \ul{bridgeno} is defined 
and is related to the adjacency matrix in App. B.

\section{Application to Gravitational Theories}
Let us apply the obtained result to some simple problems. First
the weak-field expansion of 
Riemann tensors are graphically represented as in Fig.21.
%
%
\begin{figure}
  \centerline{\epsfysize=2cm \epsfbox{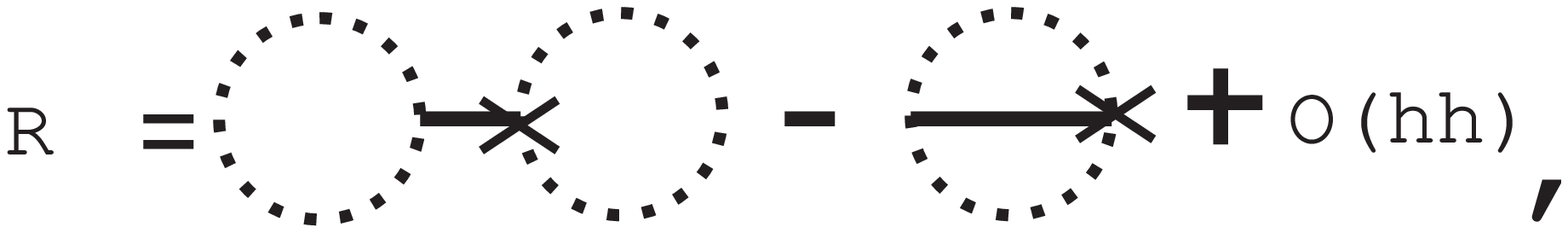}}
  \centerline{\epsfysize=7cm \epsfbox{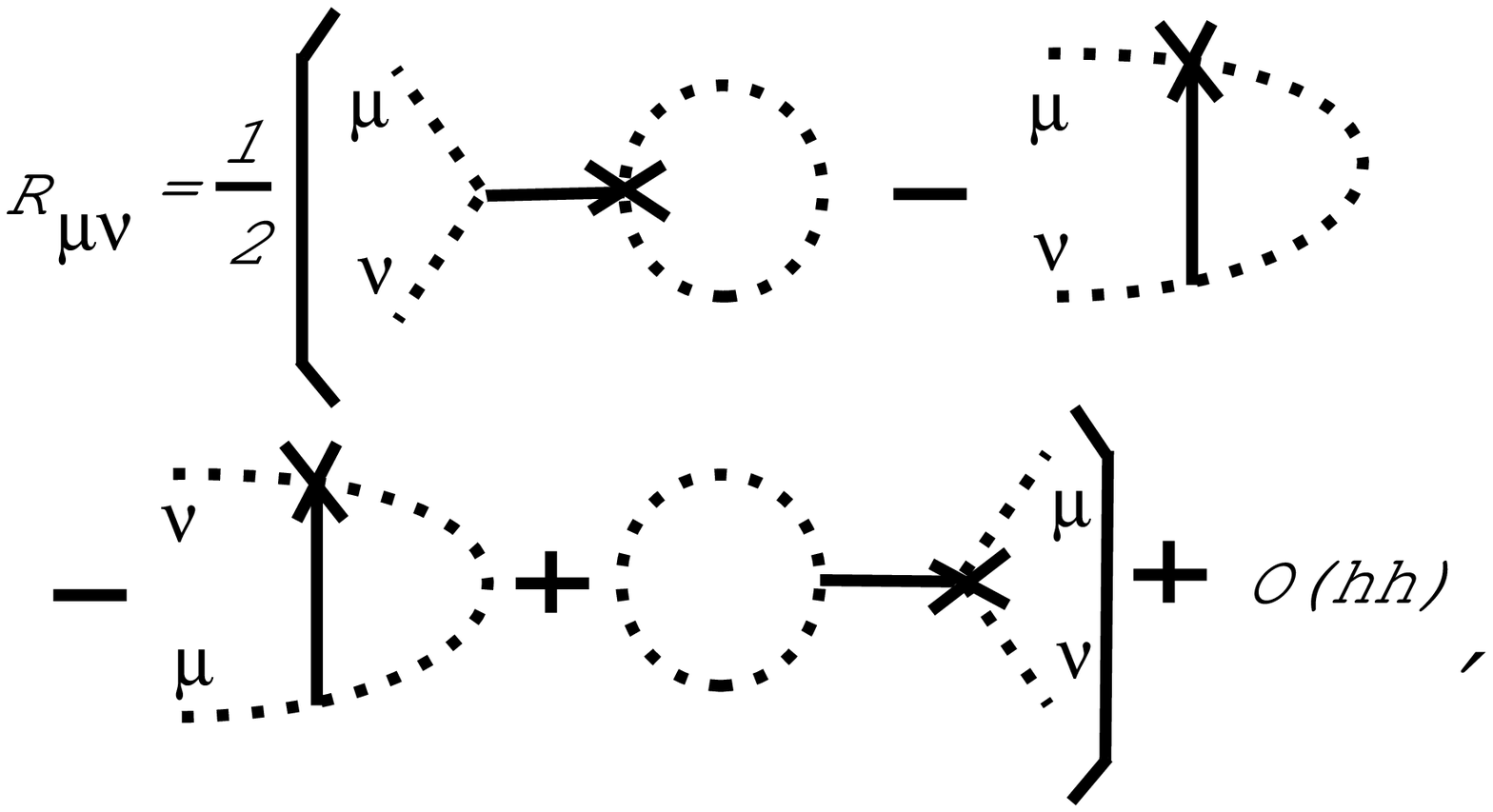}}
  \centerline{\epsfysize=5.8cm \epsfbox{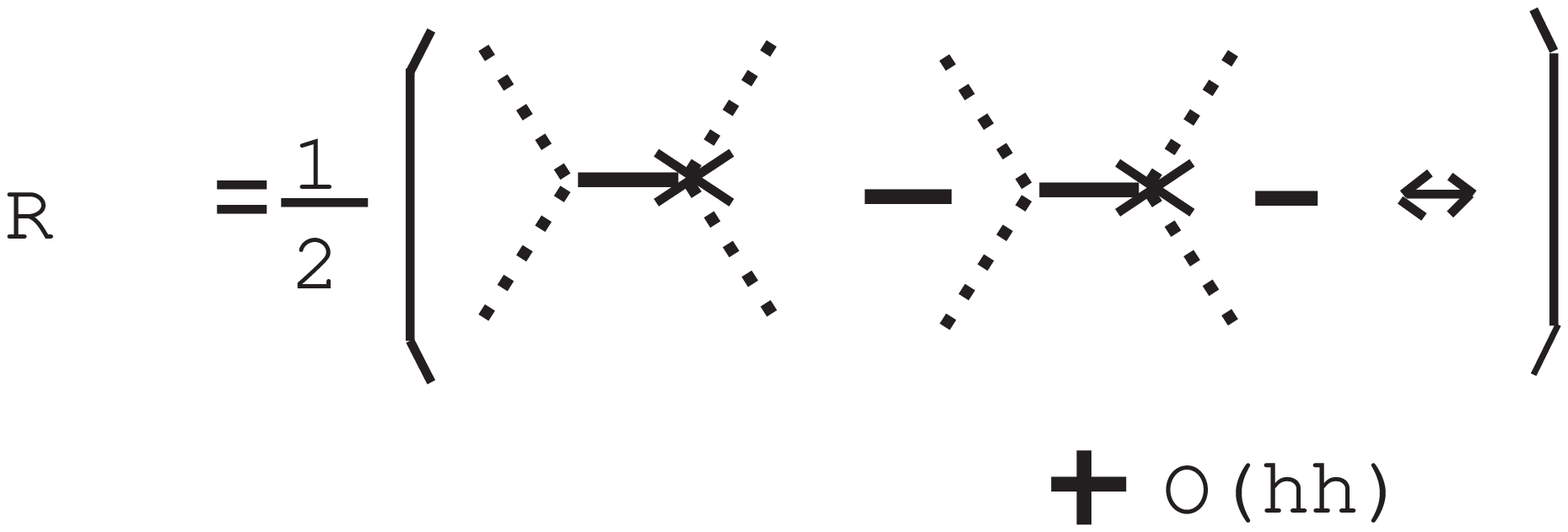}}
   \begin{center}
Fig.21\ Graphical representation of weak expansion of Riemann tensors .
   \end{center}
\end{figure}
Using them, 
general invariants with the mass dimension (Mass)$^4$ are
expanded and their $(\pl\pl h)^2$-parts are given in Table 2. 

\begin{tabular}{|c||c|c|c|c|}
\hline
Graph      & $\qq\na^2R\qq$  & $\qq R^2\qq$  & $\qq R_\mn R^\mn\qq$ 
                                             &  $R_{\mn\ls}R^{\mn\ls}$  \\
\hline
$A1 $ &    $1$       & $0$            & $0$                 & $-2$       \\
$A2 $ &    $2$       &  $0$           & $\half$             & $0$        \\
$A3 $ &    $0$       &  $0$           & $\half$             & $0$        \\
$B1 $ &    $-2$      &  $0$           & $-1$                & $0$        \\
$B2 $ &    $2$       &  $0$           & $-1$                & $0$        \\
$B3 $ &$-\frac{3}{2}$&  $0$           & $0$                 & $1$        \\
$B4 $ &    $0$       &  $0$           & $0$                 & $1$        \\
$Q^2$ &    $0$       &  $1$           & $0$                 & $0$        \\
$C1 $ & $\half$      &  $0$           & $\fourth$           & $0$        \\
$C2 $ &    $-1$      &  $0$           & $\fourth$           & $0$        \\
$C3 $ &    $-1$      &  $0$           & $\half$             & $0$        \\
$PQ $ &    $0$       &  $-2$          & $0$                 & $0$        \\
$P^2$ &    $0$       &  $1$           & $0$                 & $0$        \\
\hline
\multicolumn{5}{c}{\q}                                                 \\
\multicolumn{5}{c}{Table 2\ \  Weak-Expansion of Invariants with (Mass)$^4$-Dim.
                              :\  $(\pl\pl h)^2$-Part }\\
\end{tabular}

\vs 1

The four invariants,
$\na^2R\ ,\ R^2\ ,\ R_\mn R^\mn\ $ and $R_{\mn\ls}R^{\mn\ls}$\ , 
are important in the Weyl anomaly calculation\cite{BD,II1} 
and (1-loop) counter term
calculation in 4 dim quantum gravity\cite{tHV,IO}. 
From the explicit result of
Table 2, we see the four invariants are independent as local functions
of $h_\mn(x)$, because the 13 $(\pl\pl h)^2$-invariants are
independent each other. In particular, the three 'products' of
Riemann tensors ($R^2\ ,\ R_\mn R^\mn\ ,R_{\mn\ls}R^{\mn\ls}$)
are 'orthogonal', at the leading order of weak field perturbation,
in the space 'spanned' by the 13 $(\pl\pl h)^2$-invariants. Note here
that the independence of the four invariants is proven for a general
metric $g_\mn=\del_\mn+h_\mn$. 
As for the next higher mass dimension case, (Mass)$^6$ general
invariants, it has been shown that, in the same way as above, 
the following 17 ones are complete and independent\cite{II2}.
\begin{eqnarray}
P_1=RRR\com\ \ P_2=RR_\mn R^\mn\com\ \ P_3=RR_{\mn\ls}R^{\mn\ls}\com \nn\\
P_4=R_\mn R^{\n\la}R_\la^{~\mu}\com\ \ P_5=-R_{\mn\ls}R^{\mu\la}R^{\nu\si}
\com\ \ P_6=R_{\mn\ls}R_\tau^{~\nu\ls}R^{\mu\tau}\com\nn\\
A_1=R_{\mn\ls}R^{\si\la}_{~~~\tau\om}R^{\om\tau\n\m}\com\ \ 
B_1=R_{\mn\tau\si}R^{\n~~~\tau}_{~\la\om}R^{\la\mu\si\om}\com\nn\\
O_1=\na^\mu R\cdot \na_\mu R\com\ \ 
O_2=\na^\mu R_\ls\cdot \na_\mu R^\ls\com\nn\\ 
O_3=\na^\mu R^{\la\rho\si\tau}\cdot \na_\mu R_{\la\rho\si\tau}\com\ \ 
O_4=\na^\mu R_{\la\n}\cdot \na^\n R^\la_{~\m}\com\nn\\ 
T_1=\na^2R\cdot R\com\ \ T_2=\na^2R_\ls\cdot R^\ls\com\ \ 
T_3=\na^2R_{\la\rho\si\tau}\cdot R^{\la\rho\si\tau}\com\ \ \nn\\
T_4=\na^\m\na^\n R\cdot R_\mn\com\ \nn\\ 
S=\na^2\na^2R\pr
                               \label{appl.1}
\end{eqnarray}

 \q We consider, as the next application, the Weyl anomalies for
the gravity-matter theory in "diverse" dimensions. Anomaly
formulae are obtained in Ref.\cite{II1}. Its lowest non-trivial
order, w.r.t. the weak-field, in $n$-dim space is given by the
$t^0$-part of the trace of the following formula.
\begin{eqnarray}
G_1(x,y;t)=\int d^nz\int^t_0ds G_0(x-z;t-s)\Vvec(z)G_0(z-y;s)\com\nn\\
G_0(x;t)=\frac{1}{(4\pi t)^{n/2}}e^{-\frac{x^2}{4t}}~I_N\com 
\label{appl.2}
\end{eqnarray}
where $\Vvec(z)$ is the interaction part of the system (elliptic)
operator, $I_N$ is the $N\times N$ unit matrix (N:\ the number of 
matter-field components). $G_0(x;t)$ is the solution of the
n-dim heat equation with the temperature $t$. As the simplest model,
we take the conformal invariant gravity-scalar theory ($N=1$)
in $n$-dim space. 
\begin{eqnarray}
\Lcal=\sqg (\half \na_\m\p\na^\m\p-\frac{n-2}{8(n-1)}R\p^2)
\com\q \label{appl.3}
\end{eqnarray}
Then the operator $\Vvec(z)$ is, at the lowest order, given by
(see eq.(16) of Ref.\cite{II1})
\begin{eqnarray}
\Vvec= W_\mn\pl_\m\pl_\n+N_\m\pl_\m+M\com\nn\\
W_\mn =-h_\mn+O(h^2)\com\q  N_\la =-\pl_\m h_{\la\m}+O(h^2)\com\nn\\
M = \frac{n-2}{4(n-1)}(\pl^2h-\pl_\al\pl_\be h_\ab)
-\fourth \pl^2h+O(h^2)\pr
\label{appl.4}
\end{eqnarray}
The diagonal ($x=y$) part, $G_1(x,x;t)$, finally reduces to
\begin{eqnarray}
G_1(x,x;t)
=\frac{1}{(4\pi)^{n/2}t^{(n/2)-1}}\int d^nw\int^1_0dr G_0(w;(1-r)r)\nn\\
\times [\frac{1}{t} W_\mn(x+\sqrt{t}w)(-\frac{\del_\mn}{2r}+\frac{w_\m w_\n}{4r^2})
+\frac{1}{\sqrt{t}} N_\m(x+\sqrt{t}w)(-\frac{w_\m}{2r})+ M(x+\sqrt{t}w)]\ .
\label{appl.5}
\end{eqnarray}
Taylor-expanding $W_\mn, N_\m$ and $M$ in
the above expression w.r.t. small $t$, and the $t^0$-part 
of $G_1$ gives
the lowest order of the Weyl anomaly terms. For various dimensions,
relevant terms are graphically shown in Fig.22.
\begin{figure}
\centerline{\epsfysize=15cm\epsfbox{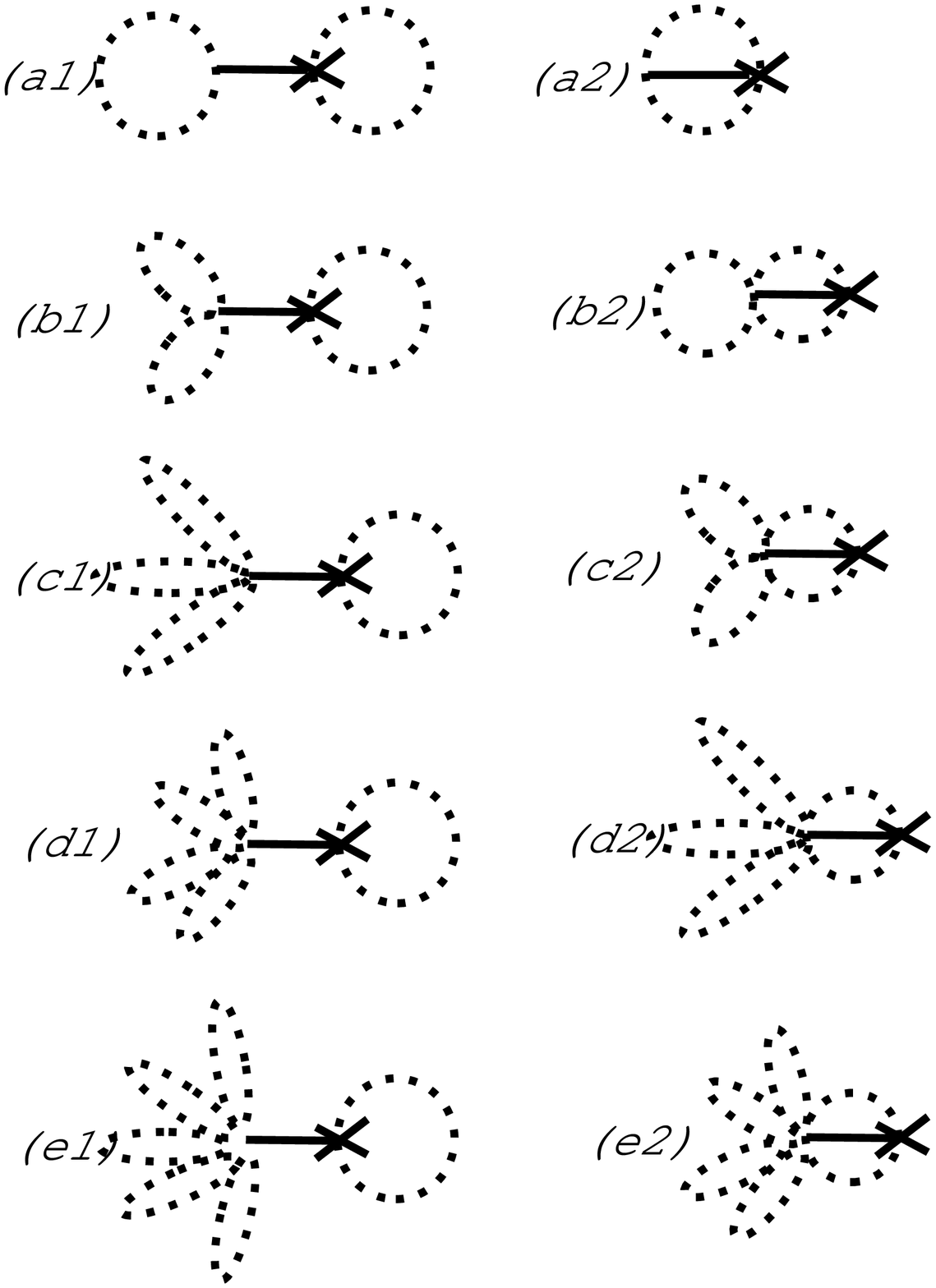}}
\begin{center}
Fig.22\ Lowest order graphs for the Weyl anomaly calculation (\ref{appl.6}):  
\nl 
[2 dim]\ (a1)\ $\pl^2h\simeq \left[ \matrix{2 & 0 \cr 0 & 2 \cr} \right]$,\ 
(a2)\ $\pl_\m\pl_\n h_\mn\simeq \left[ \matrix{0 & 2 \cr 2 & 0 \cr} \right]$; 
\nl 
[4 dim]\ (b1)\ $(\pl^2)^2h\simeq \left[ \matrix{4 & 0 \cr 0 & 2 \cr} \right]$,\ 
(b2)\ $\pl^2\pl_\m\pl_\n h_\mn\simeq \left[ \matrix{2 & 2 \cr 2 & 0 \cr} \right]$; 
\nl 
[6 dim]\ (c1)\ $(\pl^2)^3h\simeq \left[ \matrix{6 & 0 \cr 0 & 2 \cr} \right]$,\ 
(c2)\ $(\pl^2)^2\pl_\m\pl_\n h_\mn\simeq \left[ \matrix{4 & 2 \cr 2 & 0 \cr} \right]$; 
\nl 
[8 dim]\ (d1)\ $(\pl^2)^4h\simeq \left[ \matrix{8 & 0 \cr 0 & 2 \cr} \right]$,\ 
(d2)\ $(\pl^2)^3\pl_\m\pl_\n h_\mn\simeq \left[ \matrix{6 & 2 \cr 2 & 0 \cr} \right]$;
\nl 
[10 dim]\ (e1)\ $(\pl^2)^5h\simeq \left[ \matrix{10 & 0 \cr 0 & 2 \cr} \right]$,\ 
(e2)\ $(\pl^2)^4\pl_\m\pl_\n h_\mn\simeq \left[ \matrix{8 & 2 \cr 2 & 0 \cr} \right]$. 
   \end{center}
\end{figure}
The general invariant forms are obtained as follows.
\begin{eqnarray}
& \mbox{2 dim }:\ \ 
A_2=\frac{1}{4\pi}\sqg\{ -\frac{1}{6} R\}\ , & \nn\\
& \mbox{4 dim }:\ \ 
A_4=\frac{1}{(4\pi)^2}\sqg\{ -\frac{1}{180} \na^2 R+(RR-\mbox{terms})\}\ , & \nn\\
& \mbox{6 dim }:\ \ 
A_6=\frac{1}{(4\pi)^3}\sqg\{ -\frac{1}{4200} \na^4 R                 
                                  +(\na\na RR-,RRR-\mbox{terms})  \}\ , & \nn\\
& \mbox{8 dim }:\ \ 
A_8=\frac{1}{(4\pi)^4}\sqg\{ -\frac{1}{7!\cdot 2} \na^6 R   \qq\qq\qq       & \nn\\
& \qq\qq\qq          +(\na\na\na\na RR-,\na\na RRR-,RRRR-\mbox{terms})  \}\ , & \nn\\
& \mbox{10 dim }:\ \ 
A_{10}=\frac{1}{(4\pi)^5}\sqg\{ -\frac{4}{9!\cdot 33} \na^8 R  \qq\qq\qq   &  \nn\\
& +(\na\na\na\na\na\na RR-,\na\na\na\na RRR-,\na\na RRRR-,RRRRR-\mbox{terms})\}\ .&
\label{appl.6}
\end{eqnarray}
The omitted part of ($\cdots$ -terms) is given by the higher-order
calculation. The full form is well-established up to $A_4$.

\section{Conclusions and Discussions}
\q We have presented a graphical representation of global SO(n) tensors.
This approach allows us to systematically list all and independent
SO(n) invariants. 
We have proposed some new methods for it:\ the adjacency matrices, 
the Feynman diagrams, 
the topological index method, etc.. 
They all give a consistent result, but each of them has its advantageous
and disadvantageous points.
We can apply them to
any higher order invariants in principle
\footnote{
The present analysis reduces the problem to the two key points:\  
a) classification of the higher order SO(n)-invariants;\ 
b) computer program and calculation. 
The point b) is purely a technical problem. Rough estimation shows the Weyl
anomaly calculation, for example, in 10 dim is possible at the present
middle-size workstation.
}
.
As some simple examples, we have calculated all 
$\pl\pl h-,(\pl\pl h)^2-,(\pl\pl\pl h)^2-$ and $(\pl\pl h)(\pl\pl\pl\pl h)-$
invariants.  
The completeness of the list is reassured by an identity
between a combinatoric number of suffixes and weights of listed terms
due to their suffix-permutation symmetries. Some topological indices, sufficient
for discriminating 
all $\pl\pl h$- and $(\pl\pl h)^2$- invariants, are given. 
They are 
useful in practical (computer) calculation. Finally we have applied
the result to some problems in the general relativity.

\q The present graphical representation for global SO(n) tensors
is complementary to that for
general tensors given in \cite{SI}. The latter one deals with
general covariants, and its results are independent of  the perturbation.
In the general covariant representation, however, it is difficult to prove
the independence of listed general invariants because there is no independent
'bases'. On the other hand, in the present case, although the analysis 
is based on the
weak field perturbation, we have independent 'bases' (like 13 $(\pl\pl h)^2$-
invariants) at each perturbation order. It allows us to prove independence
of listed general invariants. The independence of $R^3$-type general
invariants is shown, using the present approach, in Ref.\cite{II2}.

\q Stimulated by the string theory, the importance of 
physics in the higher dimensions is increasing. Especially $n=10$\ dimensions
is the critical one. When the string physics is so well developed that
the dynamical aspect becomes clearer, we suppose that the relation
between the string field theory and the ordinary field theory becomes
seriously important. At present no consistent field heory in higher than
4 dim is known except the "free" theories like (\ref{appl.3}). We believe
the present result gives some useful tools for such analysis. For example
we must treat $(\pl\pl h)^5$-invariants, like terms graphically shown in Fig.23,
in order to completely determine $A_{10}$.  
\begin{figure}
\centerline{\epsfysize=5cm\epsfbox{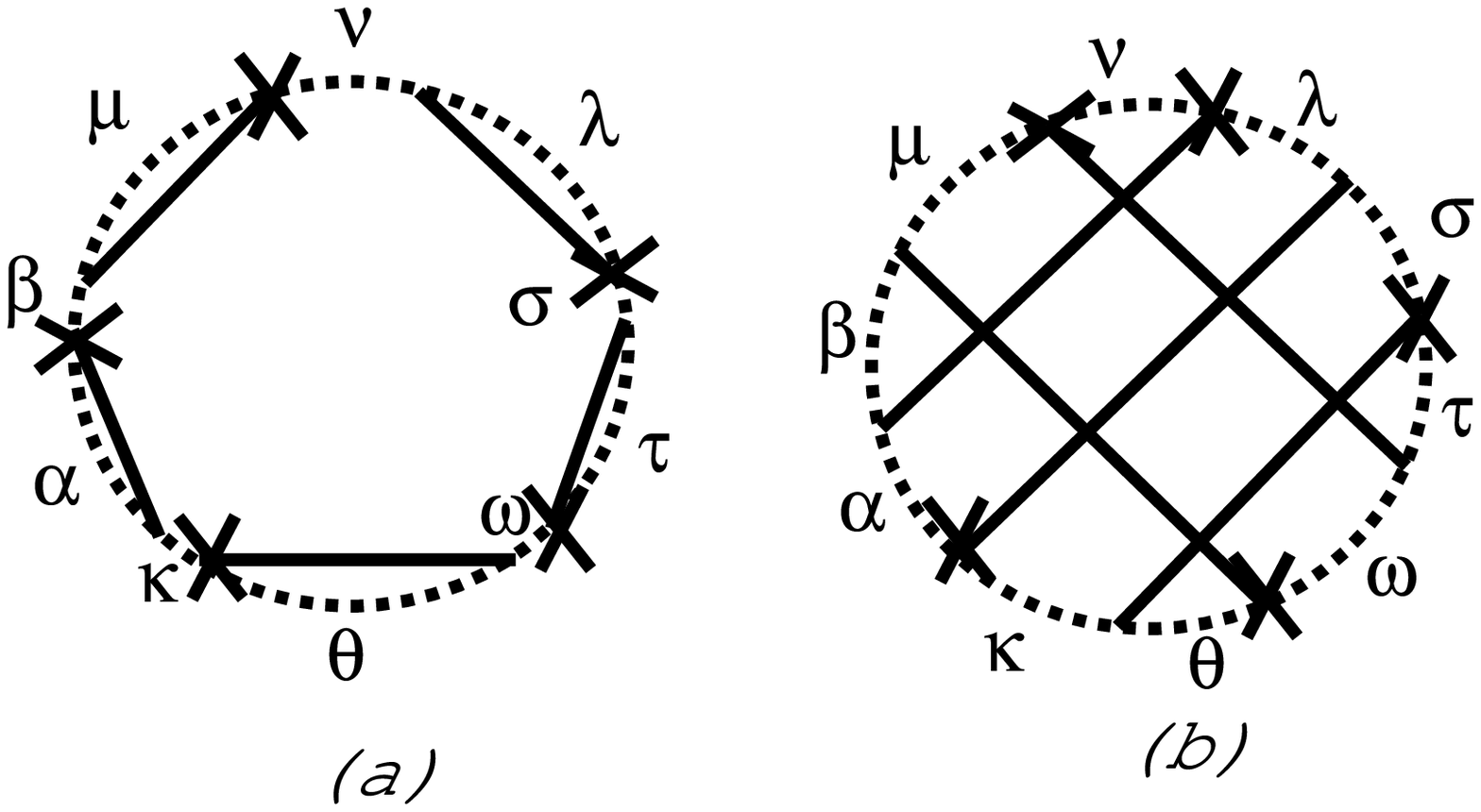}}
\begin{center}
Fig.23\ (a)\ $\pl_\al\pl_\ka h_\ab\cdot\pl_\be\pl_\m h_\mn\cdot
\pl_\n\pl_\la h_\ls\cdot \pl_\si\pl_\tau h_{\tau\om}\cdot\pl_\om\pl_\th h_{\th\ka}$  
and (b)\  
$\pl_\tau\pl_\om h_\mn\cdot\pl_\m\pl_\be h_{\om\th}\cdot\pl_\al\pl_\be h_{\n\la}
\cdot\pl_\la\pl_\si h_{\al\ka}\cdot\pl_\ka\pl_\th h_{\si\tau}$.
   \end{center}
\end{figure}

\vs 1

\q Some results such as (\ref{gc.4}) and Table 2
are obtained or checked by the computer calculation using a C-language program
\cite{SI96}.

\vs 2
\begin{flushleft}
{\bf Acknowledgement}
\end{flushleft}
The authors thank Prof.K.Murota (RIMS,Kyoto Univ.)
for discussions and comments about the present work. They express
gratitude to Prof. N.Nakanishi and Prof. K.Murota 
for reading the initial version of
the manuscript carefully. 

\vs 2
\begin{flushleft}
{\Large\bf Appendix A.\ The Young Tableaus} 
\end{flushleft}
We can calculate $SO(n)$ invariants by the Young diagram method.
The general theories are discussed in \cite{L} and \cite{FKWC}.
So we briefly explain this method taking simple examples:\ 
$\partial\partial h$- and $(\partial\partial h)^2$- invariants. 

\q The permutation symmetries w.r.t. the indices in a term, 
$\partial_{\mu_1} \cdots \partial_{\mu_d} h_{\lambda\rho}$,  
are represented by the following
Young tableaus:
\begin{center}
\setlength{\unitlength}{1.2pt}
\begin{picture}(90,20)(0,-10)
\put(0,0){\line(1,0){50}}
\put(0,-10){\line(1,0){50}}
\put(0,0){\line(0,-1){10}}
\put(10,0){\line(0,-1){10}}
\put(20,0){\line(0,-1){10}}
\put(40,0){\line(0,-1){10}}
\put(50,0){\line(0,-1){10}}
\put(25,-5){{\circle*{2}}}
\put(30,-5){{\circle*{2}}}
\put(35,-5){{\circle*{2}}}
\put(55,-5){{\circle*{2}}}
%
\put(60,0){\line(1,0){20}}
\put(60,-10){\line(1,0){20}}
\put(60,0){\line(0,-1){10}}
\put(70,0){\line(0,-1){10}}
\put(80,0){\line(0,-1){10}}
%
\end{picture},
\end{center}
where the first tableau has $d$ boxes in a row and 
"$\cdot$" is the product of the Young tableaus.
We simply express the above diagram by a symbol
arranging the box numbers of each row, so the term 
$\partial_{\mu_1} \cdots \partial_{\mu_d} h_{\lambda\rho}$ 
is expressed as 
$
\{d\}\cdot \{2\}.
$

\q $\partial_\mu \partial_\nu h_{\lambda\rho}$ is decomposed as 
\begin{eqnarray}
& \{2\}\cdot \{2\}  =  \{4\} + \{31\} + \{22\}, & \nn\\
& \left(\ 
\setlength{\unitlength}{1.2pt}
\begin{picture}(20,10)(0,0)
\put(0,5){\line(1,0){20}}
\put(0,-5){\line(1,0){20}}
\put(0,5){\line(0,-1){10}}
\put(10,5){\line(0,-1){10}}
\put(20,5){\line(0,-1){10}}
\end{picture}
\q\cdot\q
\begin{picture}(20,10)(0,0)
\put(0,5){\line(1,0){20}}
\put(0,-5){\line(1,0){20}}
\put(0,5){\line(0,-1){10}}
\put(10,5){\line(0,-1){10}}
\put(20,5){\line(0,-1){10}}
\end{picture}
\q =\q
\begin{picture}(40,10)(0,0)
\put(0,5){\line(1,0){40}}
\put(0,-5){\line(1,0){40}}
\put(0,5){\line(0,-1){10}}
\put(10,5){\line(0,-1){10}}
\put(20,5){\line(0,-1){10}}
\put(30,5){\line(0,-1){10}}
\put(40,5){\line(0,-1){10}}
\end{picture}
\q +\q
\begin{picture}(30,20)(0,0)
\put(0,10){\line(1,0){30}}
\put(0,0){\line(1,0){30}}
\put(0,-10){\line(1,0){10}}
\put(0,10){\line(0,-1){20}}
\put(10,10){\line(0,-1){20}}
\put(20,10){\line(0,-1){10}}
\put(30,10){\line(0,-1){10}}
\end{picture}
\q+\q
\begin{picture}(20,20)(0,0)
\put(0,10){\line(1,0){20}}
\put(0,0){\line(1,0){20}}
\put(0,-10){\line(1,0){20}}
\put(0,10){\line(0,-1){20}}
\put(10,10){\line(0,-1){20}}
\put(20,10){\line(0,-1){20}}
\end{picture}
\right) &
\label{ddh1}
\end{eqnarray}
by the Littlewood-Richardson rule 
to obtain irreducible representations of the symmetric groups.
The right hand side of (\ref{ddh1}) corresponds to the independent
symmetrizations or antisymmetrizations of indices in 
$\partial_\mu \partial_\nu h_{\lambda\rho}$.
We contract the indices and obtain  $SO(n)$ invariants. 
One tableau can produce one independent invariant.
The contraction, however, gives a non-zero result only when
all the box numbers of each row are even. 
Therefore in the above $\partial \partial h$ example, 
we obtain only two independent invariants from 
$\{4\}$ and $\{22\}$.
\begin{eqnarray}
\setlength{\unitlength}{1.2pt}
\begin{picture}(40,10)(0,0)
\put(0,5){\line(1,0){40}}
\put(0,-5){\line(1,0){40}}
\put(0,5){\line(0,-1){10}}
\put(10,5){\line(0,-1){10}}
\put(20,5){\line(0,-1){10}}
\put(30,5){\line(0,-1){10}}
\put(40,5){\line(0,-1){10}}
\put(2,-3){$\m$}
\put(12,-3){$\n$}
\put(22,-3){$\la$}
\put(32,-3){$\si$}
\end{picture}
\times\ \del_\mn \del_\ls\q\q
\propto\q P+Q\com\nn\\
\setlength{\unitlength}{1.2pt}
\begin{picture}(20,20)(0,0)
\put(0,10){\line(1,0){20}}
\put(0,0){\line(1,0){20}}
\put(0,-10){\line(1,0){20}}
\put(0,10){\line(0,-1){20}}
\put(10,10){\line(0,-1){20}}
\put(20,10){\line(0,-1){20}}
\put(2,2){$\m$}
\put(12,2){$\n$}
\put(2,-8){$\la$}
\put(12,-8){$\si$}
\end{picture}
\times\ \del_\mn \del_\ls\q\q
\propto\q P-Q\com
         \label{appA.3}
\end{eqnarray}
where $P\equiv\pl_\m\pl_\m h_{\n\n}$ and $Q\equiv\pl_\m\pl_\n h_{\mn}$
are defined in the text.
Generally, the independent invariants are obtained 
from the tableaus whose box numbers of each row are all even.

\q However when we calculate the independent invariants from the products 
of the {\it same} tensors, we must modify the Littlewood-Richardson rule 
to obtain the appropriate ones.
The Young tableaus to obtain the irreducible representations
of the symmetric groups does not produce the correct results in such
case.
For example, $(\partial\partial h)^2$ invariants are not the
decomposition of 
\begin{eqnarray}
(\{4\} + \{31\} + \{22\})
\cdot
(\{4\} + \{31\} + \{22\}).
\end{eqnarray}
In order to symmetrize the products,
we use the 'plethysm' operation $\otimes$,
which is explained in the appendix in \cite{L}.
Since it is the rather complicated theory, we do not explain it in
this paper.
We represent $(\partial\partial h)^2$ invariants as 
\begin{eqnarray}
(\{4\} + \{31\} + \{22\})^{\otimes \{2\}}, 
\end{eqnarray}
and they  are calculated by the plethysm method as
\begin{eqnarray}
(\{2\}\cdot \{2\})^{\otimes \{2\}} 
& = & (  \{4\} + \{31\} + \{22\} )^{\otimes \{2\}}  \\ \nonumber
& = & \{4\}^{\otimes \{2\}} + \{31\}^{\otimes \{2\}} 
+ \{22\}^{\otimes \{2\}} + \{4\}\{31\} \\ \nonumber
&& 
+ \{4\}\{22\} + \{31\}\{22\} 
\\ \nonumber
& = & \{8\} + 4 \{62\} + 3 \{44\} + 4 \{422\} + \{2222\} 
+ \mbox{uneven terms} \\ \nonumber
& = &
\setlength{\unitlength}{1.2pt}
\begin{picture}(90,30)(0,-10)
\put(0,0){\line(1,0){80}}
\put(0,-10){\line(1,0){80}}
\put(0,0){\line(0,-1){10}}
\put(10,0){\line(0,-1){10}}
\put(20,0){\line(0,-1){10}}
\put(30,0){\line(0,-1){10}}
\put(40,0){\line(0,-1){10}}
\put(50,0){\line(0,-1){10}}
\put(60,0){\line(0,-1){10}}
\put(70,0){\line(0,-1){10}}
\put(80,0){\line(0,-1){10}}
\end{picture}
+ 4 
\quad
\setlength{\unitlength}{1.2pt}
\begin{picture}(70,40)(0,-10)
\put(0,0){\line(1,0){60}}
\put(0,-10){\line(1,0){60}}
\put(0,-20){\line(1,0){20}}
\put(0,0){\line(0,-1){20}}
\put(10,0){\line(0,-1){20}}
\put(20,0){\line(0,-1){20}}
\put(30,0){\line(0,-1){10}}
\put(40,0){\line(0,-1){10}}
\put(50,0){\line(0,-1){10}}
\put(60,0){\line(0,-1){10}}
\end{picture}
+ 3
\quad
\setlength{\unitlength}{1.2pt}
\begin{picture}(50,30)(0,-10)
\put(0,0){\line(1,0){40}}
\put(0,-10){\line(1,0){40}}
\put(0,-20){\line(1,0){40}}
\put(0,0){\line(0,-1){20}}
\put(10,0){\line(0,-1){20}}
\put(20,0){\line(0,-1){20}}
\put(30,0){\line(0,-1){20}}
\put(40,0){\line(0,-1){20}}
\end{picture}
\\ \nonumber
&& 
+ 4
\quad
\setlength{\unitlength}{1.2pt}
\begin{picture}(50,40)(0,-10)
\put(0,0){\line(1,0){40}}
\put(0,-10){\line(1,0){40}}
\put(0,-20){\line(1,0){20}}
\put(0,-30){\line(1,0){20}}
\put(0,0){\line(0,-1){30}}
\put(10,0){\line(0,-1){30}}
\put(20,0){\line(0,-1){30}}
\put(30,0){\line(0,-1){10}}
\put(40,0){\line(0,-1){10}}
\end{picture}
+ 
\quad
\setlength{\unitlength}{1.2pt}
\begin{picture}(30,50)(0,-10)
\put(0,0){\line(1,0){20}}
\put(0,-10){\line(1,0){20}}
\put(0,-20){\line(1,0){20}}
\put(0,-30){\line(1,0){20}}
\put(0,-40){\line(1,0){20}}
\put(0,0){\line(0,-1){40}}
\put(10,0){\line(0,-1){40}}
\put(20,0){\line(0,-1){40}}
\end{picture}
+ \mbox{uneven terms},
\end{eqnarray}
\normalsize
\hfil\break
\hfil\break
where the numbers in front of Young tableaus above show those
of independent representations of the same type.
The 13 independent Young Tableaus give, after contraction, the 13 independent 
$(\partial \partial h)^2$-invariants, which are 
the linear combinations of 13 terms in Table 1 of the text. 

\q An advantage of this method is that 
we can find easily the relations, among invariants,  
depending on the space dimension\cite{SI}\cite{II2}.
The antisymmetricity with respect to the "vertical" boxes
in the Young tableaus require, to give a nonvanishing invariant
after the contraction, that the number of space (and time) coordinates
is larger than or equal to the maximum number of rows. 
For example, the $\{2222\}$ invariant is zero if the space dimension is
less than four, 
so that it gives one 
identity among the invariants. 
In $(\partial \partial h)^2$-invariants, we find that 
if the  dimension $n=1$, the independent invariant is one.
If $n=2$ ,there are $8$,
if $n=3$, $12$, if $n \geq 4$, $13$.

\vs 1
\begin{flushleft}
{\Large\bf Appendix B.\ Adjacency Matrices for 
$(\partial^3 h)^2$- and 
$(\partial^4 h \partial^2 h)$- 
invariants}
\end{flushleft}
\vs 1
In this appendix, we list the complete and independent 
$(\partial \partial \partial h)^2$- and 
$(\partial \partial \partial \partial h \partial \partial h)$- 
invariants.
The corresponding graphs for them are given in Ref.\cite{II2}.
We introduce an important new index by two definitions.
\begin{description}
\item[Def 9]\ 
Let us consider a general SO(n)-invariant of a binary type: 
$\pl^r h\cdot \pl^s h,\ r+s=\mbox{even}$.
(We explicitly consider the cases of $(r=3, s=3)$ and $(r=4,s=2)$.)
The invariant 
$\pl^r h\cdot \pl^s h$ is
represented by a graph
with $(r+s+4)/2$ suffix-lines where each of them
connects two vertices in the graph. We define {\it bridge-lines}
as those suffix-lines which connect a vertex of one bond
with another vertex of the other bond. 
\end{description}
\begin{description}
\item[Def 10]\ 
For a general SO(n)-invariant of a binary type: 
$\pl^r h\cdot \pl^s h,\ r+s=\mbox{even}$, we define
{\it bridge number} (\ul{bridgeno}) as the number of
bridge-lines of the graph.
\end{description}
\ul{bridgeno} must be an odd (even) number for $r=$odd (even). 
The discrimination of invariants can be done mainly by \ul{bridgeno}
and the number of suffix-loops, $\ul{l}$, which is defined in Sec.7
of the text.
We give the representatives of the adjacency matrices.
These are determined from the conditions (\ref{ad.3}) and (\ref{ad.4})
in Sec.4. The index, \ul{bridgeno}, defined above can be read from
a matrix as follows.
\begin{eqnarray}
& A=[a_{IJ}];\ I,J=1,{\bar 1},2,{\bar 2}\com  & \nn\\
& \mbox{\ul{bridgeno}}=a_{12}+a_{1{\bar 2}}+a_{{\bar 1}2}+a_{{\bar 1}{\bar 2}}
(=a_{21}+a_{2{\bar 1}}+a_{{\bar 2}1}+a_{{\bar 2}{\bar 1}})\pr &
\label{dmat.1}
\end{eqnarray}

\vs {0.5}
\begin{flushleft}
{(i)\ $(\partial^3 h)^2$ invariants}
\end{flushleft}
%
\begin{eqnarray}
4F1 = \partial^2 \partial_\mu h_{\lambda\lambda} \partial^2
\partial_\mu h_{\rho\rho} 
\simeq
\left[
\matrix{
2 & 0 & 1 & 0 \cr
0 & 2 & 0 & 0 \cr
1 & 0 & 2 & 0 \cr
0 & 0 & 0 & 2 \cr
}
\right],
3F1a = \partial^2 \partial_\mu h_{\lambda\lambda}
\partial^2 \partial_\nu h_{\mu\nu} 
\simeq
\left[
\matrix{
2 & 0 & 0 & 1 \cr
0 & 2 & 0 & 0 \cr
0 & 0 & 2 & 1 \cr
1 & 0 & 1 & 0 \cr
}
\right],
\nonumber \\
%
3F1b = \partial^2 \partial_\mu h_{\nu\nu} \partial_\mu 
\partial_\lambda \partial_\rho h_{\lambda\rho} 
\simeq
\left[
\matrix{
2 & 0 & 1 & 0 \cr
0 & 2 & 0 & 0 \cr
1 & 0 & 0 & 2 \cr
0 & 0 & 2 & 0 \cr
}
\right],
2F1a = \partial^2 \partial_\mu h_{\mu\nu}
\partial^2 \partial_\lambda h_{\lambda\nu} 
\simeq
\left[
\matrix{
2 & 1 & 0 & 0 \cr
1 & 0 & 0 & 1 \cr
0 & 0 & 2 & 1 \cr
0 & 1 & 1 & 0 \cr
}
\right],
\nonumber \\
%
3F3b = \partial^2 \partial_\mu h_{\lambda\rho} \partial^2
\partial_\mu h_{\lambda\rho} 
\simeq
\left[
\matrix{
2 & 0 & 1 & 0 \cr
0 & 0 & 0 & 2 \cr
1 & 0 & 2 & 0 \cr
0 & 2 & 0 & 0 \cr
}
\right],
2F3a = \partial^2 \partial_\mu h_{\lambda\rho}
\partial^2 \partial_\lambda h_{\mu\rho} 
\simeq
\left[
\matrix{
2 & 0 & 0 & 1 \cr
0 & 0 & 1 & 1 \cr
0 & 1 & 2 & 0 \cr
1 & 1 & 0 & 0 \cr
}
\right],
\nonumber \\
%
3F3c = \partial^2 \partial_\mu h_{\nu\lambda} \partial_\mu
\partial_\nu \partial_\lambda h_{\rho\rho} 
\simeq
\left[
\matrix{
2 & 0 & 1 & 0 \cr
0 & 0 & 2 & 0 \cr
1 & 2 & 0 & 0 \cr
0 & 0 & 0 & 2 \cr
}
\right],
3F3a = \partial_\mu \partial_\nu \partial_\lambda h_{\rho\rho}
\partial_\mu \partial_\nu \partial_\lambda h_{\sigma\sigma}
\simeq
\left[
\matrix{
0 & 0 & 3 & 0 \cr
0 & 2 & 0 & 0 \cr
3 & 0 & 0 & 0 \cr
0 & 0 & 0 & 2 \cr
}
\right],
\nonumber \\
%
2F1c = \partial^2 \partial_\mu h_{\mu\nu} \partial_\nu
\partial_\lambda \partial_\rho h_{\lambda\rho} 
\simeq
\left[
\matrix{
2 & 1 & 0 & 0 \cr
1 & 0 & 1 & 0 \cr
0 & 1 & 0 & 2 \cr
0 & 0 & 2 & 0 \cr
}
\right],
2F3b = \partial^2 \partial_\mu h_{\nu\lambda}
\partial_\nu \partial_\lambda \partial_\rho h_{\mu\rho}
\simeq
\left[
\matrix{
2 & 0 & 0 & 1 \cr
0 & 0 & 2 & 0 \cr
0 & 2 & 0 & 1 \cr
1 & 0 & 1 & 0 \cr
}
\right],
\nonumber \\
%
2F3c = \partial^2 \partial_\mu h_{\nu\lambda} \partial_\mu
\partial_\nu \partial_\rho h_{\lambda\rho} 
\simeq
\left[
\matrix{
2 & 0 & 1 & 0 \cr
0 & 0 & 1 & 1 \cr
1 & 1 & 0 & 1 \cr
0 & 1 & 1 & 0 \cr
}
\right],
3F3d = \partial_\lambda \partial_\rho \partial_\nu h_{\mu\mu}
\partial_\lambda \partial_\rho \partial_\sigma h_{\nu\sigma}
\simeq
\left[
\matrix{
0 & 0 & 2 & 1 \cr
0 & 2 & 0 & 0 \cr
2 & 0 & 0 & 1 \cr
1 & 0 & 1 & 0 \cr
}
\right],
\nonumber \\
%
2F1b = \partial_\mu \partial_\nu \partial_\sigma h_{\mu\nu}
\partial_\lambda \partial_\rho \partial_\sigma h_{\lambda\rho} 
\simeq
\left[
\matrix{
0 & 2 & 1 & 0 \cr
2 & 0 & 0 & 0 \cr
1 & 0 & 0 & 2 \cr
0 & 0 & 2 & 0 \cr
}
\right],
2F3d = \partial_\mu \partial_\nu \partial_\lambda h_{\mu\rho}
\partial_\nu \partial_\lambda \partial_\sigma h_{\rho\sigma}
\simeq
\left[
\matrix{
0 & 1 & 2 & 0 \cr
1 & 0 & 0 & 1 \cr
2 & 0 & 0 & 1 \cr
0 & 1 & 1 & 0 \cr
}
\right],
\nonumber \\
%
2F3e = \partial_\mu \partial_\nu \partial_\lambda h_{\mu\rho}
\partial_\nu \partial_\rho \partial_\sigma h_{\lambda\sigma} 
\simeq
\left[
\matrix{
0 & 1 & 1 & 1 \cr
1 & 0 & 1 & 0 \cr
1 & 1 & 0 & 1 \cr
1 & 0 & 1 & 0 \cr
}
\right],
3F5 = \partial_\mu \partial_\nu \partial_\lambda h_{\rho\sigma}
\partial_\mu \partial_\nu \partial_\lambda h_{\rho\sigma}
\simeq
\left[
\matrix{
0 & 0 & 3 & 0 \cr
0 & 0 & 0 & 2 \cr
3 & 0 & 0 & 0 \cr
0 & 2 & 0 & 0 \cr
}
\right],
\nonumber \\
%
2F5b = \partial_\mu \partial_\nu \partial_\lambda h_{\rho\sigma}
\partial_\mu \partial_\nu \partial_\rho h_{\lambda\sigma} 
\simeq
\left[
\matrix{
0 & 0 & 2 & 1 \cr
0 & 0 & 1 & 1 \cr
2 & 1 & 0 & 0 \cr
1 & 1 & 0 & 0 \cr
}
\right],
2F5a = \partial_\mu \partial_\nu \partial_\lambda h_{\rho\sigma}
\partial_\mu \partial_\rho \partial_\sigma h_{\nu\lambda}
\simeq
\left[
\matrix{
0 & 0 & 1 & 2 \cr
0 & 0 & 2 & 0 \cr
1 & 2 & 0 & 0 \cr
2 & 2 & 0 & 0 \cr
}
\right],\nn\\
\label{dddhdddh.1}
\end{eqnarray}
The naming of invariants follows the rule:\ the first number shows that
of the suffix-loops ($\ul{l}$), the last number shows that of the
bridge number ($\ul{bridgeno}$). It is the same for the next case (ii).
The lowest order of $\na R\times \na R$-invariants are given by the linear
combination of the above listed terms.
\begin{eqnarray}
O_1=\na^\mu R\cdot \na_\mu R=(2F1b)-2(3F1b)+(4F1)\com\nn\\ 
O_2=\na^\mu R_\ls\cdot \na_\mu R^\ls
=-(2F3c)-(3F3d)+\half \{ (2F3d)+(2F3e)+(3F3c) \}      \nn\\
                             +\fourth \{ (3F3a)+(3F3b)\}  \com\nn\\ 
O_3=\na^\mu R^{\la\rho\si\tau}\cdot \na_\mu R_{\la\rho\si\tau}
=(2F5a)-2(2F5b)+(3F5)                                  \com\nn\\ 
O_4=\na^\mu R_{\la\n}\cdot \na^\n R^\la_{~\m}
=-(3F3d)+\half\{ -(2F3b)-(2F3c)+(3F3c) \}               \nn\\
+\fourth\{ (2F3a)+3(2F3d)+(2F3e)+(3F3a) \} \com 
                               \label{dddhdddh.2}
\end{eqnarray}

\begin{flushleft}
{(ii) $(\partial^4 h \partial^2 h)$-invariants}
\end{flushleft}

\begin{eqnarray}
P'P = \partial^2 \partial^2 h_{\lambda\lambda} \partial^2
h_{\rho\rho} 
\simeq
\left[
\matrix{
4 & 0 & 0 & 0 \cr
0 & 2 & 0 & 0 \cr
0 & 0 & 2 & 0 \cr
0 & 0 & 0 & 2 \cr
}
\right],
P'Q = \partial^2 \partial^2 h_{\lambda\lambda}
\partial_\mu \partial_\nu h_{\mu\nu} 
\simeq
\left[
\matrix{
4 & 0 & 0 & 0 \cr
0 & 2 & 0 & 0 \cr
0 & 0 & 0 & 2 \cr
0 & 0 & 2 & 0 \cr
}
\right],
\nonumber \\
%
4H2b = \partial^2 \partial^2 h_{\lambda\rho} \partial^2
h_{\lambda\rho} 
\simeq
\left[
\matrix{
4 & 0 & 0 & 0 \cr
0 & 0 & 0 & 2 \cr
0 & 0 & 2 & 0 \cr
0 & 2 & 0 & 0 \cr
}
\right],
4H2a = \partial^2 \partial^2 h_{\mu\nu}
\partial_\mu \partial_\nu h_{\lambda\lambda} 
\simeq
\left[
\matrix{
4 & 0 & 0 & 0 \cr
0 & 0 & 2 & 0 \cr
0 & 2 & 0 & 0 \cr
0 & 0 & 0 & 2 \cr
}
\right],
\nonumber \\
%
3H2a = \partial^2 \partial^2 h_{\mu\nu} \partial_\mu \partial_\lambda
h_{\nu\lambda} 
\simeq
\left[
\matrix{
4 & 0 & 0 & 0 \cr
0 & 0 & 1 & 1 \cr
0 & 1 & 0 & 1 \cr
0 & 1 & 1 & 0 \cr
}
\right],
Q'P = \partial^2 \partial_\mu \partial_\nu h_{\mu\nu}
\partial^2 h_{\lambda\lambda} 
\simeq
\left[
\matrix{
2 & 0 & 0 & 0 \cr
0 & 2 & 0 & 0 \cr
0 & 0 & 2 & 2 \cr
0 & 0 & 2 & 0 \cr
}
\right],
\nonumber \\
%
4H2d = \partial^2 \partial_\mu \partial_\nu h_{\lambda\lambda} 
\partial^2 h_{\mu\nu} 
\simeq
\left[
\matrix{
2 & 0 & 0 & 2 \cr
0 & 2 & 0 & 0 \cr
0 & 0 & 2 & 0 \cr
2 & 0 & 0 & 0 \cr
}
\right],
4H2c = \partial^2 \partial_\mu \partial_\nu h_{\lambda\lambda} 
\partial_\mu \partial_\nu
h_{\rho\rho} 
\simeq
\left[
\matrix{
2 & 0 & 2 & 0 \cr
0 & 2 & 0 & 0 \cr
2 & 0 & 0 & 0 \cr
0 & 0 & 0 & 2 \cr
}
\right],
\nonumber \\
%
3H2b = \partial^2 \partial_\mu \partial_\nu h_{\lambda\lambda}
\partial_\mu \partial_\rho h_{\nu\rho} 
\simeq
\left[
\matrix{
2 & 0 & 1 & 1 \cr
0 & 2 & 0 & 0 \cr
1 & 0 & 0 & 1 \cr
1 & 0 & 1 & 0 \cr
}
\right],
3H2d = \partial^2 \partial_\mu \partial_\nu h_{\mu\lambda} 
\partial^2 h_{\nu\lambda} 
\simeq
\left[
\matrix{
2 & 1 & 0 & 1 \cr
1 & 0 & 0 & 1 \cr
0 & 0 & 2 & 0 \cr
1 & 1 & 0 & 0 \cr
}
\right],
\nonumber \\
%
3H2c = \partial^2 \partial_\mu \partial_\nu h_{\mu\lambda}
\partial_\nu \partial_\lambda h_{\rho\rho} 
\simeq
\left[
\matrix{
2 & 1 & 1 & 0 \cr
1 & 0 & 1 & 0 \cr
1 & 1 & 0 & 0 \cr
0 & 0 & 0 & 2 \cr
}
\right],
Q'Q = \partial^2 \partial_\mu \partial_\nu h_{\mu\nu} 
\partial_\lambda \partial_\rho h_{\lambda\rho} 
\simeq
\left[
\matrix{
2 & 2 & 0 & 0 \cr
2 & 0 & 0 & 0 \cr
0 & 0 & 0 & 2 \cr
0 & 0 & 2 & 0 \cr
}
\right],
\nonumber \\
%
2H2b = \partial^2 \partial_\mu \partial_\nu h_{\mu\lambda}
\partial_\nu \partial_\rho h_{\lambda\rho} 
\simeq
\left[
\matrix{
2 & 1 & 1 & 0 \cr
1 & 0 & 0 & 1 \cr
1 & 0 & 0 & 1 \cr
0 & 1 & 1 & 0 \cr
}
\right],
2H2a = \partial^2 \partial_\mu \partial_\nu h_{\mu\lambda} 
\partial_\lambda \partial_\rho h_{\nu\rho} 
\simeq
\left[
\matrix{
2 & 1 & 0 & 1 \cr
1 & 0 & 1 & 0 \cr
0 & 1 & 0 & 1 \cr
1 & 0 & 1 & 0 \cr
}
\right],
\nonumber \\
%
3H4b = \partial^2 \partial_\mu \partial_\nu h_{\lambda\rho}
\partial_\mu \partial_\nu h_{\lambda\rho} 
\simeq
\left[
\matrix{
2 & 0 & 2 & 0 \cr
0 & 0 & 0 & 2 \cr
2 & 0 & 0 & 0 \cr
0 & 2 & 0 & 0 \cr
}
\right],
3H4c = \partial^2 \partial_\mu \partial_\nu h_{\lambda\rho} 
\partial_\lambda \partial_\rho h_{\mu\nu} 
\simeq
\left[
\matrix{
2 & 0 & 0 & 2 \cr
0 & 0 & 2 & 0 \cr
0 & 2 & 0 & 0 \cr
2 & 0 & 0 & 0 \cr
}
\right],
\nonumber \\
%
2H4c = \partial^2 \partial_\mu \partial_\nu h_{\lambda\rho}
\partial_\mu \partial_\lambda h_{\nu\rho} 
\simeq
\left[
\matrix{
2 & 0 & 1 & 1 \cr
0 & 0 & 1 & 1 \cr
1 & 1 & 0 & 0 \cr
1 & 1 & 0 & 0 \cr
}
\right],
3H2f = \partial_\mu \partial_\nu \partial_\lambda \partial_\rho
h_{\lambda\rho} 
\partial^2 h_{\mu\nu} 
\simeq
\left[
\matrix{
2 & 0 & 0 & 0 \cr
0 & 0 & 2 & 0 \cr
0 & 2 & 0 & 2 \cr
0 & 0 & 2 & 0 \cr
}
\right],
\nonumber \\
%
3H2e = \partial_\mu \partial_\nu \partial_\rho \partial_\sigma
h_{\rho\sigma}
\partial_\mu \partial_\nu h_{\lambda\lambda} 
\simeq
\left[
\matrix{
0 & 0 & 2 & 0 \cr
0 & 2 & 0 & 0 \cr
2 & 0 & 0 & 2 \cr
0 & 0 & 2 & 0 \cr
}
\right],
3H4a = \partial_\mu \partial_\nu \partial_\lambda \partial_\rho
h_{\sigma\sigma} 
\partial_\mu \partial_\nu h_{\lambda\rho} 
\simeq
\left[
\matrix{
0 & 0 & 2 & 2 \cr
0 & 2 & 0 & 0 \cr
2 & 0 & 0 & 0 \cr
2 & 0 & 0 & 0 \cr
}
\right],
\nonumber \\
%
2H2c = \partial_\mu \partial_\nu \partial_\lambda \partial_\rho
h_{\mu\nu}
\partial_\lambda \partial_\sigma h_{\rho\sigma} 
\simeq
\left[
\matrix{
0 & 2 & 1 & 1 \cr
2 & 0 & 0 & 0 \cr
1 & 0 & 0 & 1 \cr
1 & 0 & 1 & 0 \cr
}
\right],
2H4a = \partial_\mu \partial_\nu \partial_\lambda \partial_\rho
h_{\mu\sigma} 
\partial_\nu \partial_\lambda h_{\rho\sigma} 
\simeq
\left[
\matrix{
0 & 1 & 2 & 1 \cr
1 & 0 & 0 & 1 \cr
2 & 0 & 0 & 0 \cr
1 & 1 & 0 & 0 \cr
}
\right],
\nonumber \\
%
2H4b = \partial_\mu \partial_\nu \partial_\lambda \partial_\rho
h_{\mu\sigma}
\partial_\nu \partial_\sigma h_{\lambda\rho} 
\simeq
\left[
\matrix{
0 & 1 & 1 & 2 \cr
1 & 0 & 1 & 0 \cr
1 & 1 & 0 & 0 \cr
2 & 0 & 0 & 0 \cr
}
\right],
\label{ddddhddh.1}
\end{eqnarray}
The lowest order of $\na\na R\times R$-invariants are given by the linear
combination of the above listed terms.
\begin{eqnarray}
T_1=\na^2R\cdot R=(Q'Q)-(Q'P)-(P'Q)+(P'P)    \com\nn\\ 
T_2=\na^2R_\ls\cdot R^\ls
=\half \{ (2H2a)+(2H2b)-(3H2a)-(3H2b)-(3H2c)-(3H2d) \}  \nn\\
+\fourth\{ (4H2a)+(4H2b)+(4H2c)+(4H2d) \}    \com\nn\\ 
T_3=\na^2R_{\la\rho\si\tau}\cdot R^{\la\rho\si\tau}
=-2(2H4c)+(3H4b)+(3H4c)                       \com\nn\\
T_4=\na^\m\na^\n R\cdot R_\mn=(2H2c)-(3H2b)      \nn\\
+\half\{ -(3H2e)-(3H2f)+(4H2c)+(4H2d) \}
                                              \pr 
                               \label{ddddhddh.2}
\end{eqnarray}

\begin{flushleft}
{\Large\bf Appendix C.\ Calculation of \ul{bcn}[\ ] and \ul{vcn}[\ ]
}
\end{flushleft}
\vs 1

\q We explain how to calculate the indices, \ul{bcn}[\ ] and \ul{vcn}[\ ] 
in the actual(computer) calculation. Let us consider a
$(\pl\pl h)^2$-invariant.
It has two bonds.
As an example, we take C1 in Fig.24.

\begin{description}
\item[Def 11]\ 
We assign i=0 for one bond and i=1 for the other. 'i' is the 
{\it bond number} and discriminates the two bonds. Next we assign j=0
for all dd-vertices and j=1 for all h-vertices. 
'j' is the {\it vertex-type number}
and discriminate the  vertex-type. Any vertex in a graph is specified
by a pair (i,j). 
\end{description}
\begin{figure}
\centerline{\epsfysize=5cm\epsfbox{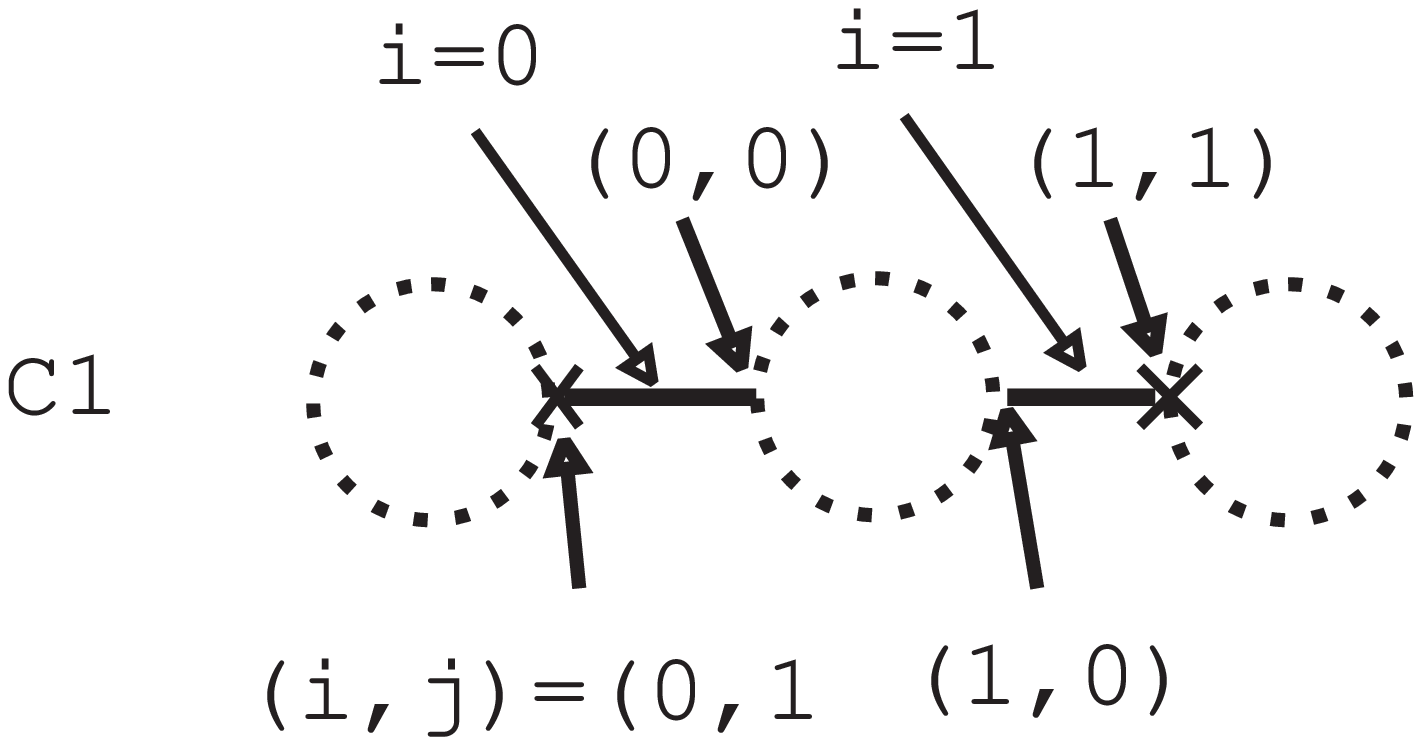}}
   \begin{center}
Fig.24\ 
Bond number 'i' and vertex-type number 'j' for each vertex \nl in 
the invariant C1.
   \end{center}
\end{figure}
\begin{description}
\item[Def 12]\ 
When we trace a suffix-line, along a loop, starting from a vertex 
(i$_0$,j$_0$) in a certain direction, we pass some vertices,
(i$_1$,j$_1$),(i$_2$,j$_2$),$\cdots$ and finally come back to the
starting vertex (i$_0$,j$_0$). 
We focus on the {\it change} of the bond number, i, and the
vertex-type number, j, when we pass from a vertex to the next vertex in 
the tracing (see Fig.25 and 18).
For $k$-th loop, we assign as
$\sum_{\mbox{along $k$-th loop}}|\Delta \mbox{i}|\equiv$ \ul{bcn}[$k$],
$\sum_{\mbox{along $k$-th loop}}|\Delta \mbox{j}|\equiv$ \ul{vcn}[$k$].
\end{description}

\begin{figure}
\centerline{\epsfysize=12cm\epsfbox{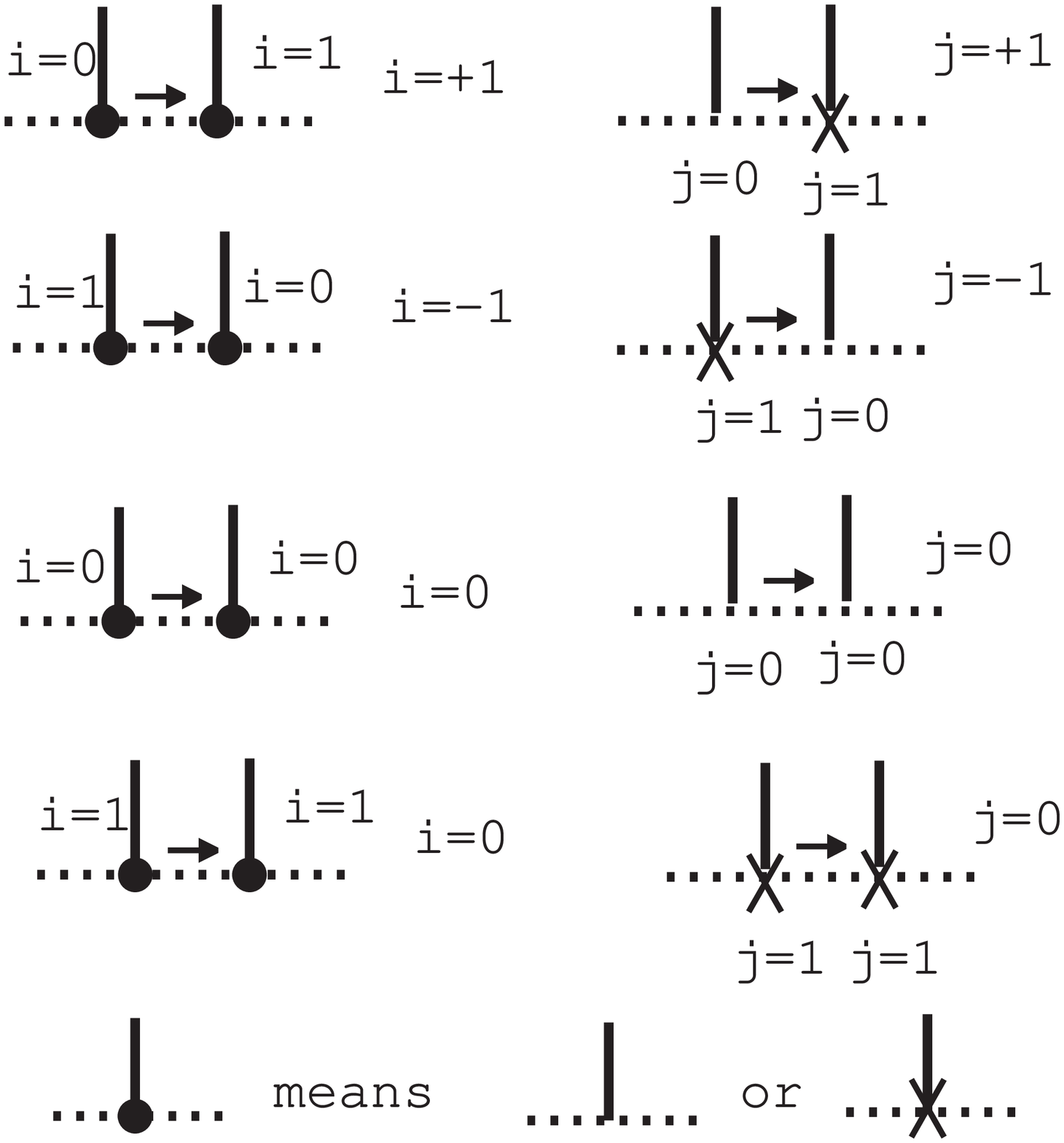}}
   \begin{center}
Fig. 25\ Change of i (bond number) and j (vertex-type number).\nl
Arrows indicate directions of tracings.
   \end{center}
\end{figure}

\ul{bcn}[\ ] and \ul{vcn}[\ ] are listed for all $(\pl\pl h)^2$-invariants in Table 1.
\ul{bcn}[\ ] and \ul{vcn}[\ ] defined above satisfy the following
important properties.
\begin{enumerate}
\item
They donot depend on the starting vertex for tracing along a loop.
\item
They donot depend on the direction of the tracing.
\end{enumerate}
\begin{figure}
\centerline{\epsfysize=4cm\epsfbox{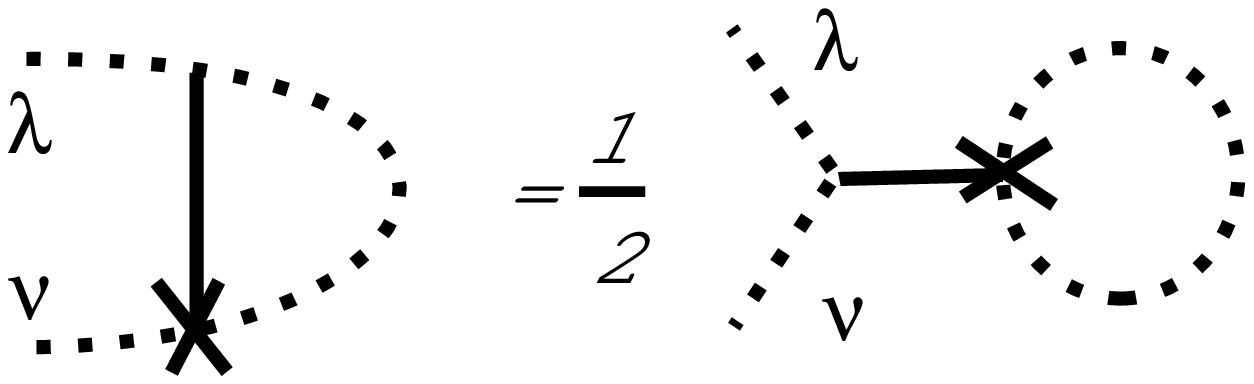}}
   \begin{center}
Fig.26\ Graphical rule, expressing  (\ref{gf.2}), 
due to the gauge-fixing condition (\ref{gf.1}) .
   \end{center}
\end{figure}

\vs 2
\begin{flushleft}
{\Large\bf Appendix D.\ 
Gauge-Fixing Condition and Graphical Rule
}
\end{flushleft}
\vs 1

\q In the text, we have not taken a gauge-fixing condition.
When we calculate a physical quantity
in the classical and quantum gravity, we sometimes need to impose the condition
on the metric $g_\mn$ for some reasons.  Firstly, in the case of quantizing
gravity itself or of solving a classical field equation with respect to
the gravity mode, 
we {\it must} impose the fixing condition in order to eliminate
the local freedom ($\ep^\m(x), \m=1,2,\cdots,n-1,n.$)
due to the general coordinate invariance (\ref{intro.2}):
$g_\mn\ra g_\mn+g_{\m\la}\na_\n\ep^\la+g_{\n\la}\na_\m\ep^\la$. 
Secondly, even when the condition is theoretically not necessary ( such as
the  quantization on the fixed curved space, or the ordinary anomaly
calculation), the gauge-fixing is practically useful because it 
considerably reduces
the number of SO(n) invariants to be considered.

\q In the weak gravity case
$g_\mn=\del_\mn+h_\mn\ ,\ |h_\mn|\ll 1$, the condition is expressed
by $h_\mn$. Let us take a familiar gauge:
\begin{eqnarray}
\pl_\m h_\mn=\half\pl_\n h\com\q h\equiv h_{\la\la}\pr\q   \label{gf.1}
\end{eqnarray}
This condition leads to the following condition on the present
basic element $\pl_\m\pl_\n h_\ab$ .
\begin{eqnarray}
\pl_\la\pl_\m h_\mn=\half\pl_\la\pl_\n h\com\q h\equiv h_{\la\la}\pr\q
                                                               \label{gf.2}
\end{eqnarray}
This gives us a graphical rule shown in Fig.26.

\q Let us see how does this rule reduce the number of independent
invariants given in the text. 
For $\pl\pl h$-invariants, we obtain the following
relation
\begin{eqnarray}
Q=\half P\pr   \label{gf.3}
\end{eqnarray}
For $(\pl\pl h)^2$-invariants, we obtain the following relations.
\begin{eqnarray}
A2=A3=\half B1=\frac{1}{4} C1\com\nn\\
B2=\half C3\com\q QQ=\half PQ=\frac{1}{4}PP\pr      \label{gf.4}
\end{eqnarray}
Therefore, in the gauge (\ref{gf.1}), we can reduce the number of independent
invariants from 2 to 1 for $\pl\pl h$-invariants (,say, $P$) and from 13 to 7
for $(\pl\pl h)^2$-invariants (,say, $A1,B3,B4,C1,C2,C3,PP$). 

\q We expect
this gauge-fixed treatment is practically very useful when 
the quantity under consideration is known 
to be gauge-invariant in advance.

\newpage
\begin{flushleft}
{\Large\bf Figure Captions}
\end{flushleft}
\begin{itemize}
\item
Fig.1
\ (a)\ (k+2)-tensor (\ref{Rep.3});\ (b)\ 4-tensor $\pl_\m\pl_\n h_\ab$
\item
Fig.2
\ 2-tensors of 
$\pl^2h_\ab\ ,\ \pl_\m\pl_\n h_{\al\al}\ $ and $\pl_\m\pl_\be h_\ab\ $
\item
Fig.3\ Invariants of 
$P\equiv \pl_\m\pl_\m h_{\al\al}$\ and $Q\equiv \pl_\al\pl_\be h_{\ab}\ $.
\item
Fig.4\ Feynman rule of the Lagrangian (\ref{conc.1}).
\item
Fig.5\ Graph of (\ref{Rep.1}).
\item
Fig.6\ Two vertices are connected by $k$ dotted lines. $\bullet$-vertex
represents a der-vertex or a h-vertex.
\item
Fig.7\ A vertex is connected with itself by $l$ dotted lines. 
$\bullet$-vertex represents a der-vertex or a h-vertex.
\item
Fig.8\ 
Graphical Representations of 
$\pl_\m\pl_\n h_\ab \pl_\m\pl_\n h_{\ga\del}$ and
$\pl_\m\pl_\n h_\ab \pl_\n\pl_\la h_{\la\be}$.
\item
Fig.9\ 
Two ways to place two dd-vertices ( small circles) and two h-vertices
(cross marks) upon one suffix-loop.
\item
Fig.10\ 
Three independent $(\pl\pl h)^2$-invariants for the case of one suffix-loop.
\item
Fig.11\ 
Bondless diagrams for (\ref{ddh2.3}).
\item
Fig.12\ 
Five independent $(\pl\pl h)^2$-invariants for the case of two suffix-loops.
\item
Fig.13\ 
Three bondless diagrams corresponding to (\ref{ddh2.4}).
\item
Fig.14\ 
Four independent $(\pl\pl h)^2$-invariants for the case of three suffix-loops.
\item
Fig.15\ 
The bondless diagram corresponding to (\ref{ddh2.5}).
\item
Fig.16\ 
The unique independent $(\pl\pl h)^2$-invariant for the case of four suffix-loops.
\item
Fig.17\ 
Graph B1 for the weight calculation (\ref{gc.3}).
\item
Fig.18\ Explanation of \ul{bcn}[\ ] and \ul{vcn}[\ ] using Graph A2.
\item
Fig.19\ Graph G9$\equiv \pl_\m\pl_\om h_\mn\cdot\pl_\n\pl_\la h_{\tau\si}
\cdot\pl_\la\pl_\si h_{\tau\om}$.
\item
Fig.20\ Graph of G25$\equiv \pl_\m\pl_\n h_\ls\cdot\pl_\si\pl_\tau h_\mn
\cdot\pl_\la\pl_\om h_{\tau\om}$.
\item
Fig.21\ Graphical representation of weak expansion of Riemann tensors .
\item
Fig.22\ Lowest order graphs for the Weyl anomaly calculation (\ref{appl.6}):  
\nl 
[2 dim]\ (a1)\ $\pl^2h\simeq \left[ \matrix{2 & 0 \cr 0 & 2 \cr} \right]$,\ 
(a2)\ $\pl_\m\pl_\n h_\mn\simeq \left[ \matrix{0 & 2 \cr 2 & 0 \cr} \right]$; 
\nl 
[4 dim]\ (b1)\ $(\pl^2)^2h\simeq \left[ \matrix{4 & 0 \cr 0 & 2 \cr} \right]$,\ 
(b2)\ $\pl^2\pl_\m\pl_\n h_\mn\simeq \left[ \matrix{2 & 2 \cr 2 & 0 \cr} \right]$; 
\nl 
[6 dim]\ (c1)\ $(\pl^2)^3h\simeq \left[ \matrix{6 & 0 \cr 0 & 2 \cr} \right]$,\ 
(c2)\ $(\pl^2)^2\pl_\m\pl_\n h_\mn\simeq \left[ \matrix{4 & 2 \cr 2 & 0 \cr} \right]$; 
\nl 
[8 dim]\ (d1)\ $(\pl^2)^4h\simeq \left[ \matrix{8 & 0 \cr 0 & 2 \cr} \right]$,\ 
(d2)\ $(\pl^2)^3\pl_\m\pl_\n h_\mn\simeq \left[ \matrix{6 & 2 \cr 2 & 0 \cr} \right]$;
\nl 
[10 dim]\ (e1)\ $(\pl^2)^5h\simeq \left[ \matrix{10 & 0 \cr 0 & 2 \cr} \right]$,\ 
(e2)\ $(\pl^2)^4\pl_\m\pl_\n h_\mn\simeq \left[ \matrix{8 & 2 \cr 2 & 0 \cr} \right]$. 
\item
Fig.23\ (a)\ $\pl_\al\pl_\ka h_\ab\cdot\pl_\be\pl_\m h_\mn\cdot
\pl_\n\pl_\la h_\ls\cdot \pl_\si\pl_\tau h_{\tau\om}\cdot\pl_\om\pl_\th h_{\th\ka}$  
and (b)\  
$\pl_\tau\pl_\om h_\mn\cdot\pl_\m\pl_\be h_{\om\th}\cdot\pl_\al\pl_\be h_{\n\la}
\cdot\pl_\la\pl_\si h_{\al\ka}\cdot\pl_\ka\pl_\th h_{\si\tau}$.
\item
Fig.24\ 
Bond number 'i' and vertex-type number 'j' for each vertex in 
the invariant C1.
\item
Fig. 25\ Change of i (bond number) and j (vertex-type number).\nl
Arrows indicate directions of tracings.
\item
Fig.26\ Graphical rule, expressing  (\ref{gf.2}), 
due to the gauge-fixing condition (\ref{gf.1}) .
\end{itemize}

\end{document}